\documentclass[twoside,a4paper]{article}
\usepackage{amsmath,graphicx,amssymb,fancyhdr,amsthm,,textcomp,hyperref}

\usepackage[english]{babel}
\usepackage[numbers]{natbib}
\usepackage[acronym, nopostdot,nomain,nonumberlist,numberedsection]{glossaries}

\usepackage{mathtools}
\usepackage{tensor}
\makenoidxglossaries

\usepackage{amsmath,graphicx,amssymb,fancyhdr,amsthm,enumerate,textcomp}

\usepackage[usenames]{color}
\usepackage{enumitem}
\newtheorem{thm}{Theorem}

\newtheorem{prop}[thm]{Proposition}

\def\sp{{\text{sp}}}
\def\SS{\mathbf{S}}

\def\l{\lambda}
\def\Dtil{\tilde{\nabla}}
\def\diag{\textrm{diag}}
\def\Gu{G_{\bf u}}

\def\I{{\bf I}}
\def\II{{\bf II}}
\def\III{{\bf III}}
\def\N{{\bf N}}
\def\D{{\bf D}}
\def\O{{\bf O}}
\def\S{{\cal S}}
\def\E{{\cal E}}
\def\Snul{{\cal S}^{r}_0}
\def\Spl{{\cal S}^{r}_+}
\def\HS{H_\S}
\def\ss{\mathfrak{s}}
\def\s{\sigma}
\def\sp{s^*}
\def\tp{t^*}
\def\Hp{H^*}
\def\pp{\mathfrak{p}^*}
\def\pf{\mathfrak{p}}

\def\IS{I_S}
\def\JS{J_S}

\usepackage{color}

\newcommand\commentout[1]{} 
\usepackage{soul} 

\def\beq{\begin{eqnarray}}
\def\eeq{\end{eqnarray}}
\def\beqs{\begin{eqnarray*}}
\def\eeqs{\end{eqnarray*}}

\numberwithin{equation}{section}

\begin{document}


\title{Observer-based invariants 
for cosmological models}

\author{
	{\large L.\ Wylleman$^\dagger$, A.\ Coley$^\diamond$, D.\ McNutt$^\dagger$ and M.\ Aadne$^\dagger$}\\
	\vspace{0.05cm} \\
	{\small $^\dagger$ Faculty of Science and Technology, University of Stavanger},\\ {\small  N-4036 Stavanger, Norway}\\
	{\small $^\diamond$ Department of Mathematics and Statistics, Dalhousie University}, \\ {\small Halifax, Nova Scotia, Canada B3H 3J5}\\
	{\small E-mail: \texttt{lode.wylleman@uis.no,aac@mathstat.dal.ca,}} \\ 
	\hspace{17 mm} {\small \texttt{david.d.mcnutt@uis.no, matthew.t.aadne@uis.no}} 
}


\date{}  
\maketitle

 
\begin{abstract}
	
	We consider the equivalence problem for cosmological models in four-dimensional gravity theories. A cosmological model is {considered} as a triple $(M, {\bf g}, {\bf u})$ consisting of a 
	spacetime $(M, {\bf g})$
	and a preferred normalized time-like vector field ${\bf u}$ tangent to a congruence of fundamental observers. We introduce a modification of the Cartan-Karlhede algorithm by restricting to frames adapted to ${\bf u}$ and including the covariant derivatives of ${\bf u}$ 
	along with the Riemann tensor and its covariant derivatives. To fix the frame we make use of quantities {\em relative to the fundamental observers}, such as the anisotropic pressure tensor, energy flux vector, electric and magnetic parts of the Weyl tensor and the kinematical quantities of {\bf u}.   
	This 
	provides a simpler way to construct a 
	list of invariants relative to the fundamental observers
	that completely characterizes the model, independent of coordinates. 
	As an illustration of the algorithm, we consider several well-known cosmological models from General Relativity.

\end{abstract}

\section{Introduction}

One of the most successful and useful applications of Einstein's theory of General Relativity (GR) has been within the field of cosmology. The first model of the universe arising from GR that matches current observations was the spatial homogeneous and isotropic model, consisting of  the so called Friedmann-Lema\^{i}tre-Robertson-Walker (FLRW) cosmologies. As observing the Universe has become more sophisticated, cosmologists have continued to produce new and interesting exact models. There are three necessary ingredients to appropriately model our Universe at a particular scale. We must choose a metric, $g_{ab}$, which characterizes the spacetime geometry representing the universe on some specific averaged scale \cite{ellis1999}. The matter content of the universe must be chosen next on the same averaged scale in terms of a stress-energy tensor $T_{ab}$, which also necessitates the existence of a preferred timelike vector field {\bf u} as an intrinsic property (see 
\cite{ellis1999} and below). Finally, the interaction of the geometry with the matter must be determined through some governing equation, such as Einstein's field equations for GR:
\beq 
R_{ab} - \frac12 R g_{ab} = T_{ab} - \Lambda g_{ab}  \label{EFE}
\eeq
\noindent where $R_{ab}$ is the Ricci tensor of the geometry, $R$ the Ricci scalar, and $\Lambda$ the cosmological constant. It is assumed that at large scales the dynamics of the spacetime geometry is governed by GR but other  gravitational theories are permitted as well. We will call such a model of the universe {\it a cosmological model}. Due to the variety of cosmological models and the approaches to generate them (see \cite{ellis1999} and below), it is of interest to determine when two cosmological models are equivalent. 

In this paper we will treat a cosmological model as a triple, $(M, {\bf g}, {\bf u})$, consisting of a four-dimensional (4D) 
spacetime, $(M, {\bf g})$, and a normalized timelike vector field, ${\bf u}$. This is motivated by the observation that for cosmological models, the existence of a unique locally defined timelike congruence with tangent field {\bf u} (which can, in principle, be extended to the whole manifold), representing a family of fundamental observers is guaranteed. This congruence is usually associated with the 4-velocity of the averaged matter in the model; i.e., the matter admits a formulation in terms of an averaged matter content which defines an average (macroscopic) timelike congruence~\cite{ellis1999}. 
As an example, for an orthogonal Bianchi, single fluid model {\bf u} is the normalized 4-velocity vector field of the fluid, i.e., the normalized timelike eigenvector of the Ricci tensor at each point. If there are additional matter sources providing further macroscopic timelike congruences, then we will seek out and identify a fundamental macroscopic timelike congruence. 
For all cosmological models there is always one timelike vector field which has a physical meaning. 

The metric ${\bf g}$ of a cosmological model can be written as a line element $ds^2=g_{\alpha\beta} dx^\alpha dx^\beta$ in a coordinate system $(x^\alpha)$, where the metric components $g_{\alpha\beta}$ contain some free functions which should be matched to obervational data. The problem of equivalence now manifests itself due to the freedom of coordinates. For instance, it is possible that one or more free functions can be `gauged away' (set to a normalized value, like 0 or 1) by making a simple coordinate transformation such as a rescaling of  coordinates or a translation of the origin of the coordinate system; then such scalars do not have a physical meaning. Also, a specific approach to generate a cosmological model comes with a `natural' coordinate system for the line element; therefore, the metrics of two models that are generated by different approaches can look completely different but may yet be equivalent. In general, two metrics with components $g_{\alpha\beta}$ and $\tilde{g}_{\mu\nu}$ in respective coordinate systems $(x^\alpha)$ and $(y^{\mu})$ (where $\alpha,\,\beta,\,\mu$ and $\nu$ run from 0 to 3) are equivalent if and only if a coordinate transformation $y^\mu=y^\mu(x^\alpha)$ (in shorthand, $y=y(x)$) exists such that 
\begin{equation}\label{g-equiv}
g_{\alpha\beta}(x)=\tilde{g}_{\mu\nu}(y(x))\frac{\partial y^\mu}{\partial x^\alpha}(x)\frac{\partial y^\nu}{\partial x^\beta}(x).
\end{equation}
It may be extremely difficult to decide in a direct way whether such a coordinate transformation exists or not. 
Note that even in the case where the 4-velocity ${\bf u}={\bf \tilde{u}}$ is fixed and a preferred timelike coordinate $x^0\equiv t=\tilde{t}\equiv y^0$ exists such that two seemingly different metrics can be written as $ds^2=-dt^2+g_{ij} dx^i dx^j$ and $d\tilde{s}^2=-dt^2+\tilde{g}_{mn}dy^mdy^n$ (where $i,\,j,\,m$ and $n$ run from 1 to 3), a coordinate transformation (or diffeomorphism) $y^m=y^m(x^i)$ may exist mapping the former to the latter; however, it remains hard to decide on existence in a direct way. 
 
Hence one would like to utilize a more convenient method to decide on equivalence. For this purpose it is crucial to note that if \eqref{g-equiv} holds then all scalar {\em invariants} which can be calculated from the two metrics (and which, unlike functions which appear in metric components but can possibly be gauged away, determine the true physical content of the models) will be the same -- they will only be expressed in different coordinates. It is therefore natural to ask whether a set of scalar invariants exists which {\em completely characterizes} a given cosmological model and can be constructed by means of a systematic procedure. Here a complete characterization means that if any other model is to be compared to the given model, one can construct the set of scalar invariants using the procedure for both metrics, in the respective coordinates $x^\alpha$ and $y^\mu$, and the two models are equivalent precisely when a coordinate transformation $y=y(x)$ exists which maps corresponding scalars to one another. Note that this is considerably simpler than to decide whether a transformation $y=y(x)$ exists which realizes \eqref{g-equiv}. Optimally, we would like the procedure to be fully algorithmic and equally applicable to all cosmological models, such that it can easily be implemented as a computer program.

Probably the most natural scalar invariants are the {\em scalar polynomial curvature invariants} (SPIs), which are full contractions of the Riemann tensor and its covariant derivatives (jointly called {\em curvature tensors} henceforth). However, it is known that the set consisting of all SPIs,
\beqs \mathcal{I} = \{ R, R_{ab}R^{ab}, C_{abcd}C^{abcd},\hdots, R_{abcd;e} R^{abcd;e},\hdots, R_{abcd;ef} R^{abcd;ef},\hdots \}, \label{Iset} \eeqs 
\noindent 
cannot be used to uniquely characterize {\em all} spacetimes. 
Consider a smooth one-parameter deformation  $\tilde{\bf g}_\tau,\,0\leq \tau\leq \epsilon$ of a Lorentzian metric ${\bf g}$, where $\tilde{\bf g}_0 = {\bf g}$ and $\tilde{\bf g}_\tau$ is not diffeomorphic to ${\bf g}$ for $\tau>0$~\cite{ColHerPel09}. If for any such deformation the set $\mathcal{I}$ for $\tilde{\bf g}_\epsilon$ differs from the original set $\mathcal{I}$ for ${\bf g}$ then the metric is called {\em $\mathcal{I}$-non-degenerate} and the metric is characterized uniquely by its SPIs. Conversely, if some deformation leaves the set $\mathcal{I}$ unchanged then the spacetime is $\mathcal{I}$-degenerate and cannot be uniquely characterized by SPIs. It is known that in 4D the $\mathcal{I}$-degenerate spacetimes consist of degenerate Kundt spacetimes~\cite{ColHerPel09,Kundt}. However, degenerate Kundt spacetimes are not viable as cosmological models.\footnote{There are $\mathcal{I}$-degenerate spacetimes that are relevant to cosmology, such as the maximally symmetric spacetimes (Minkowski and (anti) de Sitter spacetimes) or the homogeneous plane wave solutions of Bianchi type IV, VI$_h$, VII$_h$ as some well-known examples. However, these solutions describe the asymptotic properties of cosmological models \cite{ siklos1991stability,Hervik:2004qv} and are not cosmological models  themselves.} 
Hence we will assume by definition that the spacetime geometry of a cosmological model is $\mathcal{I}$-non-degenerate. 

While cosmological models can thus be characterized uniquely by their SPIs, another problem arises. Computationally it is necessary to determine a minimal basis for the SPIs constructed from the Riemann tensor and its covariant derivatives up to some order. Although a basis of SPIs formed from the Riemann tensor alone is known \cite{CM1991}, no such basis is known for the higher order SPIs. In addition, the actual calculation of SPIs can be computationally difficult, which can further obstruct the use of SPIs for spacetime classification. 


An alternative set of invariants which completely characterizes {\em any} spacetime is provided by the {\em Cartan-Karlhede algorithm}. The invariants in question are simply the components of the curvature tensors relative to a preferred frame of some particular type (e.g.\ of orthonormal or complex null type) and are called {\em Cartan invariants}. To determine a preferred frame the Cartan-Karlhede algorithm employs canonical forms of the curvature tensors. This is achieved in the first instance by using the canonical form of the Weyl tensor or Ricci tensor in accord with their algebraic structure; i.e., by using the Petrov type or Segre type for the Weyl and Ricci tensors, respectively (see Chapters 4 and 5 of \cite{Stephani}, and section \ref{sec:CKalg} and \ref{subsec:CKcosm-types} below). The present implementation of the algorithm puts the Weyl tensor in a canonical form first, and 
then uses the residual frame freedom to normalize the Ricci tensor.

While this procedure is technically an algorithm, it sometimes requires input from the user to choose which components of the higher order curvature tensors to normalize, due to the absence of a canonical form for these tensors. 
However, if a spacetime admits additional invariant structure, this can be exploited to simplify the choice of frame. In the case of cosmological models, an orthonormal frame can be constructed using the additionally given timelike vector field, ${\bf u}$, giving a covariant $1\,+\,3$ split of
spacetime, and consequently the set of permitted Lorentz frame transformations to further fix the frame is considerably smaller: 
\begin{prop} The proper Lorentz transformations that preserve {\bf u} are the spatial rotations in the 3-dimensional transverse space ${\bf u}^\bot$. 
\end{prop} 

With this observation we will introduce an algorithm to uniquely characterize (locally) a cosmological model $(M, {\bf g}, {\bf u})$. 
The proposed scheme will be simpler to implement than the standard algorithm for at least two reasons.\footnote{However, we emphasize that our scheme is not a substitute for the standard algorithm, since for geometries $(M, {\bf g})$ a preferred timelike vector field ${\bf u}$ is not given.} First, it 
works with observer-based quantities (relative Riemann quantities, in the first instance; see section \ref{sec:prelim}) which are defined in the transverse spaces of the fundamental observers and can be represented by scalars and 3D vectors and symmetric tensors; these are easier to deal with than the tensors used in the 4D algorithm and moreover have a more transparent interpretation. Second, the higher derivatives of the curvature tensor will be supplemented by covariant derivatives of ${\bf u}$; in particular, the scheme also uses the kinematical quantities of {\bf u} (representing its first covariant derivative).  By doing so one constructs a list of extended Cartan invariants relative to the fundamental observers which completely characterizes the model in a coordinate independent manner. In addition to classification, such an invariant characterization of cosmological models contributes to their physical interpretation. 

The present article is organized as follows. In section 2, we fix the notation and recall basic definitions needed for this work. 
In section 3, we review the Cartan-Karlhede algorithm for spacetimes. In section 4, we introduce and discuss our modified algorithm for cosmological models. In section 5, we apply the modified algorithm to five cosmological models in GR as an illustration. In section 6, we provide a summary and a brief final discussion. 

\section{Preliminary}\label{sec:prelim}

\noindent {\em General notation and conventions.} $(M,{\bf g})$ denotes a sufficiently regular spacetime; the metric ${\bf g}$ has signature sequence $(-+++)$.
For tensor fields (in short: tensors) on $M$ we use either boldfaced, index-free notation, or abstract index notation with small Latin letters a, b,\ldots. The norm of a vector ${\bf v}$ is denoted $|{\bf v}|$. The (inverse) metric can be used to lower (raise) indices, leading to geometrically equivalent tensors denoted by the same symbols; e.g.,\ $v_a=g_{ab}v^b$. A tensor with $m$ indices is called an $m$-tensor. Round (square) brackets denote (anti)symme\-trization. The orthogonal complement $\{x^a|x^a v_a=0\}$ of a vector ${\bf v}$ is denoted ${\bf v}^\bot$. The Levi-Civita covariant derivative is denoted $\nabla$; associated are the totally antisymmetric tensor $\eta_{abcd}=\eta_{[abcd]}$, Riemann tensor $R_{abcd}$, Ricci tensor $R_{ab}=R^c_{~acb}$, Ricci scalar $R=R^a_{~a}$, and Weyl tensor $C_{abcd}$ defined by\footnote{The Weyl tensor has the same symmetries as the Riemann tensor and is trace-free 
($C_{(ab)cd}=C_{ab(cd)}=C_{a[bcd]}=0,\,C_{abcd}=C_{cdab},\,C^a{}_{bac}=0$).} 
\begin{equation}
\label{Riedecomp}
R^{ab}{}_{cd}=
C^{ab}{}_{cd}+2\delta^{[a}_{[c}R^{b]}{}_{d]}-\frac{1}{3}R\,\delta^a_{[c}\delta^b_{d]}.
\end{equation}
 
\noindent {\em Unit timelike vectors.} A unit timelike vector field $u^a$ ($g_{ab}u^au^b=-1$) on $(M,{\bf g})$ results in a triple $(M,{\bf g},{\bf u})$. One defines 
\beq
h_{ab}=g_{ab}+u_au_b,\qquad \eta_{abc}=\eta_{abcd}u^d.  
\eeq 
The tensor $h^a{}_b$ projects vectors at a spacetime point onto the instantaneous rest space (transverse space) ${\bf u}^\bot$ of an observer moving with 4-velocity $u^a$, and $\eta_{abc}$ serves as a volume element on ${\bf u}^\bot$:
$$
h^a_{~c} h^c_{~b} = h^a_{~b},~~h^a_{~a} = 3,~~h^a_{~b} u^b = 0,\qquad \eta_{abc}=\eta_{[abc]},~~\eta_{abc}u^c=0. 
$$  

An orthonormal (ON) frame is denoted $(e_0^a,e_i^a)$, where $i$ labels the spacelike vectors $e_i^a$ and runs from 1 to 3. An ON frame is called {\em {\bf u}-adapted} when $e_0^a=u^a$. 

Angle brackets will denote the orthogonally projected part of vectors and the orthogonally projected symmetric trace-free (PSTF) part of 2-tensors~\cite{Maa1997,ellis1999}: 
\begin{equation}\label{def-angle}
v^{\langle a\rangle}=h^a_{~b}v^b,\qquad S_{\langle ab\rangle} 
= \left(h^c{}_{(a} h^d{}_{b)} - \tfrac13 h_{ab} h^{cd}\right)\, S_{cd}.
\end{equation}

\noindent {\em Spatial vectors and PSTF 2-tensors.}
A vector ${\bf v}\in {\bf u}^\bot$  ($\Leftrightarrow v^au_a=0\Leftrightarrow v^a=v^{\langle a\rangle}$) is called {\em spatial} (relative to {\bf u}). Taking any ON triad $(e_i^a)$ of ${\bf u}^\bot$ we can represent {\bf v} by its $1\times 3$ vector of components relative to the triad. 

A 2-tensor $\SS= S_{ab}$ is called PSTF (relative to {\bf u}) if 
$$S_{ab}=S_{\langle ab\rangle}\quad\Leftrightarrow\quad S^a_{~a} = 0,\;\;S_{ab} = S_{(ab)},\;\;S_{ab} u^b = 0.$$ 
Given any ON triad $(e_i^a)$ of ${\bf u}^\bot$, {\bf S} can be represented by a $3\times 3$ trace-free symmetric matrix consisting of its spatial components. 
The associated endomorphism of ${\bf u}^\bot$ which transforms $v^a$ to $S^a{}_b v^b$ 
will also be denoted $\SS$. It has real eigenvalues $\l_i$ summing to zero:
\begin{equation}\label{lambdasumzero}
\l_1+\l_2+\l_3=0.
\end{equation}
Moreover, it allows for an ON eigentriad $(e_i^a)$ in which the representation matrix of $\SS$ takes a diagonal form: $\SS\equiv \diag(\l_1,\l_2,\l_3)$. The eigenvalues solve the characteristic equation $x^3-\tfrac12 I_S x-\tfrac13 J_S=0$, where 
\begin{align}
&\IS=S^a{}_bS^b{}_a=\l_1^2+\l_2^2+\l_3^2,\label{def-IS}\\ 
&\JS=S^a{}_bS^b{}_cS^c{}_a=\l_1^3+\l_2^3+\l_3^3=3\l_1\l_2\l_3\label{def-JS}
\end{align} 
are the quadratic and cubic trace invariants of $\SS$. The last equality of \eqref{def-JS} is due to \eqref{lambdasumzero}, which also implies that there are only two possibilities:
\begin{itemize}
	\item $\IS^3\neq 6\JS^2$. Then $\SS$ has distinct eigenvalues $\l_1\neq\l_2\neq\l_3\neq\l_1$.
	\item $\IS^3=6\JS^2\neq 0$. In this case $\SS$ has a double eigenvalue $\l=-\JS/\IS$ with a corresponding 2D eigenplane, and a single eigenvalue $-2\l$ with a unique eigendirection orthogonal to the eigenplane. 
\end{itemize}

\noindent {\em Derivatives.} Related to {\bf u} there are two types of covariant derivatives acting on arbitrary tensors 
$M^{a_1\cdots}{}_{b_1\cdots}$~\cite{ellis1999}: 
the covariant `time' derivative along 
{\bf u}, denoted 
\beq\label{deriv-time}
\dot{M}^{a\cdots}{}_{b\cdots}=u^c\,\nabla_c {M}^{a\cdots}{}_{b\cdots},
\eeq 
and the fully orthogonally projected covariant derivative, denoted by
\beq\label{deriv-proj}
\Dtil_c {M}^{a\cdots}{}_{b\cdots} = h_c^{~f} h^a_{~d}\cdots h^e_{~b}\cdots \nabla_{f} T^{d\cdots}{}_{e\cdots}. 
\eeq

\noindent {\em Relative Riemann and kinematical quantities.} 
Due to Einstein's field equations \eqref{EFE} the Ricci tensor is equivalent to the energy-momentum tensor: 
\beq\label{Ttensor} 
T_{ab} = R_{ab}-\tfrac{1}{2}R g_{ab}+\Lambda g_{ab}\quad\Leftrightarrow\quad R_{ab} = T_{ab}-\tfrac{1}{2}T g_{ab}+\Lambda g_{ab}, 
\eeq
where $\Lambda$ is the cosmological constant and $T=T^a{}_a=4\Lambda-R$. Relative to {\bf u} one defines the relativistic energy density $\mu$, isotropic pressure $p$, tracefree anisotropic pressure  $\pi_{ab}$, and  momentum density or energy flux $q^a$, by 
\begin{align}
\label{rel-Ric-mup}
&\mu = T_{ab} u^a u^b=G_{ab}u^au^b-\Lambda,\quad p= \frac13 T_{ab}h^{ab}=\frac13 G_{ab}h^{ab}+\Lambda,\\
\label{rel-Ric-piq} 
&\pi_{ab}=T_{\langle ab\rangle}=G_{\langle ab\rangle},\quad q^a=- T^{\langle a\rangle }{}_{b} u^b=- G^{\langle a\rangle }{}_{b} u^b,
\end{align}
with $G_{ab}=R_{ab}-\frac12 Rg_{ab}$ the Einstein tensor, and the electric and magnetic Weyl tensors by 
\beq 
\label{rel-Weyl}
E_{ab} = C_{acbd} u^c u^d,\quad H_{ab} = \tfrac12 \eta_{aef} C^{ef}{}_{bd}\,u^d.
\eeq
We call \eqref{rel-Ric-mup}-\eqref{rel-Weyl} the {\em relative Riemann quantities}; they completely determine $T_{ab}$ and $C_{abcd}$, and thus the Riemann tensor, according to\footnote{The sum of the first two terms in \eqref{rel-Weyl}, involving $E_{ab}$, is the {\em electric part} of $C^{ab}{}_{cd}$; the remaining part is the {\em magnetic part} of $C^{ab}{}_{cd}$~\cite{HerOrtWyl13}.}
\begin{align}
\label{T-decomp}
&T_{ab} = \mu u_a u_b + p h_{ab} + 2q_{(a} u_{b)}  + \pi_{ab}\\
\label{C-decomp}
&C^{ab}{}_{cd}= 4\left(u^{[a}u_{[c}+h^{[a}{}_{[c}\right)E^{b]}{}_{d]}+2\left(\eta^{abe}u_{[c}H_{d]e}+\eta_{cde}u^{[a}H^{b]e}\right).
\end{align}

Regarding the covariant derivative $\nabla_b u_a$ one defines the {\em kinematical quantities} of {\bf u}, that is, the rate of volume expansion scalar $\Theta$, rate of shear tensor $\s_{ab}$, vorticity vector $\omega^a$,\footnote{The Hubble scalar is $\Theta/3$, while $\omega^a$ and $\s_{ab}$ describe the rotation of the matter relative to a Fermi-propagated frame \cite{ellis1999} and the rate of distortion of the matter flow, respectively.} and acceleration vector $\dot{u}^a$, by
\begin{equation}
\label{kin-u}
\Theta=\nabla_a u^a,\quad \s_{ab}=\nabla_{\langle a}u_{b\rangle},\quad \omega^a=\tfrac12\eta^{abc}\nabla_c u_b,\quad \dot{u}^a=u^b\nabla_b u^a.
\end{equation}
These completely represent $\nabla_b u_a$ according to 
\beq 
\nabla_b u_a = - \dot{u}_a u_b  + \Dtil_b u_a,\qquad 
\Dtil_b u_a= \eta_{abc}\omega^c+ \frac13 \Theta h_{ab} + \sigma_{ab}. 
\label{uderiv} 
\eeq

\noindent {\em Isotropy group and LRS property of tensors.} Consider a finite set or list ${\cal S}$ of tensors and a point $p\in M$. 
For simplicity we assume that the indices of each tensor have been lowered. Given a basis $B=(e_\alpha^a)$ of $T_p M$, an $m$-tensor ${\bf S}\in {\cal S}$ can be represented by an array $V_{\bf S}$
of its components $S_{\alpha_1\cdots \alpha_m}=S_{a_1\cdots a_m}e_{\alpha_1}^{a_1}\cdots e_{\alpha_m}^{a_m}$ relative to $B$, where $\alpha_1,\ldots \alpha_m$ run from 1 to 4. Let $G$ be the Lorentz group acting on $T_p M$ (consisting of the linear transformations that preserve ${\bf g}$-inner products of vectors). If $g\in G$ transforms $B$ to the basis $B'=(M^\beta{}_\alpha e_{\beta}^a)$, then the components $M^{\beta_1}{}_{\alpha_1}\cdots M^{\beta_m}{}_{\alpha_m} S_{\beta_1\cdots \beta_m}$ of {\bf S} relative to $B'$ form a new array $V'_{\bf S}$. Doing this for all tensors in ${\cal S}$ we obtain arrays $V_{\cal S}$ and $V'_{\cal S}$. By definition, $g$ is said to belong to the isotropy group of ${\cal S}$ (or to `leave ${\cal S}$ invariant') at $p$ if $V_{\cal S}=V'_{\cal S}$.
This definition makes sense since it is independent of the chosen basis $B$, as is easily shown; moreover, in the case where ${\cal S}$ is a list it is independent of the ordering of the elements. The tensors we will be working with (essentially, the curvature tensors and the fundamental unit timelike vector field)
are assumed to be sufficiently regular so that the isotropy groups of ${\cal S}$ at each point $p$ of (any neighbourhood of) $M$ are all conjugate, and thus have the same dimension; cf.\ \cite{Stephani}. This defines a section of the bundle $M\times G\rightarrow M$ to which we refer as the {\em isotropy group}
of ${\cal S}$, denoted $H_{\cal S}$. If $H_{\cal S}$ has a one-dimensional subgroup which at each point consists of the rotations in some spacelike 2D plane, then ${\cal S}$ is called {\em locally rotationally symmetric} (LRS) in this plane.

\section{The Cartan-Karlhede method for determining local equivalence of spacetimes}\label{sec:CKalg}

The Cartan-Karlhede method provides a unique local characterization of sufficiently 
regular spacetime geometries $(M,{\bf g})$. The method uses the Riemann curvature tensor and its covariant derivatives where the consecutive steps in the procedure correspond to successive orders of covariant derivation. The main idea is to reduce the bundle of frames of some particular type (e.g.,\ null or orthonormal frames) as much as possible at each order by casting the curvature tensors up to that order into a canonical form, and at the next order only permitting those frame transformations which preserve this canonical form. 

The computer implementation in four spacetime dimensions often works with null frames (or tetrads),  $(k^a, l^a, m^a {\bar m^a})$ such that $k^ak_a=l^a l_a=m^am_a= {\bar m^a}{\bar m_a}=0$ and $-k^a l_a=1=m^a {\bar m_a}$ and where a bar denotes complex conjugation. { In our context 
it will be more convenient to use orthonormal (ON) frames 
$(e_0^a,e_i^a)$,
which can be bijectively related to null frames by} 
\beq\label{ON-null}
k^a=\tfrac{1}{\sqrt{2}}(e_0^a+e_1^a),\quad l^a=\tfrac{1}{\sqrt{2}}(e_0^a-e_1^a),\quad m^a=\tfrac{1}{\sqrt{2}}(e_2^a-ie_3^a). 
\eeq
In terms of the dual coframe $(\Omega^0_a,\Omega^i_a)$ given by  $\Omega^0_a=-(e_0)_a=-g_{ab}e_0^b$ and $\Omega^i_a=(e_i)_a=g_{ab}e_i^b$ the line element is diagonal:
\beq\label{inverse-ortho}
ds^2=-(\Omega^0)^2+(\Omega^1)^2+(\Omega^2)^2+(\Omega^3)^2.
\eeq

Working with ON frames, the Cartan-Karlhede algorithm may be summarized as follows  \cite{Mac86}:

\begin{enumerate}
	\item Let $H_{-1}$ be the full Lorentz group $G$, $s_{-1}=\dim(H_{-1})=6$ and $t_{-1}=0$, and set the order of differentiation $q$ to 0.
	\item Calculate the curvature tensor of the $q$th order: $\nabla_{e_q}\cdots\nabla_{e_1}R_{abcd}$.
	\item Use the group $H_{q-1}$ to normalize this tensor, thus finding a canonical form of the list of the curvature tensors up to the $qth$ order,
	\begin{equation}
	{\cal S}_q=(R_{abcd};\,\ldots\,;\nabla_{e_q}\cdots\nabla_{e_1}R_{abcd}), 
	\end{equation}
	and determine $H_q=H_{{\cal S}_q}$ and $s_q=\dim(H_q)$, the dimension of the remaining vertical part of the frame bundle. 
	\item Find the number $t_q$ of independent functions of spacetime position in the components of the elements of ${\cal S}_q$, 
	in the canonical form; $4-t_q$ gives the remaining horizontal freedom.
	\item If $s_q=s_{q-1}$ and $t_q=t_{q-1}$ then 
	let $\pf+1=q$ and stop the algorithm; if this is not the case then 
	increase $q$ by 1 and go to step 2. 
\end{enumerate}

\noindent The numbers $s_q$ and $t_q$ do not depend on the canonical forms used \cite{Olverbook,MilWyl13}.
For a certain choice of canonical forms the components of the curvature tensors (up to order $q$) are referred to as ($q$th order) {\em Cartan invariants}. A statement of the minimal set of Cartan invariants required, taking Bianchi and Ricci identities into account, was given in \cite{MacAma86}. We will refer to 
invariants constructed from, or equal to, Cartan invariants of any order  
as {\it extended invariants}. 
Thus, for sufficiently 
regular metrics, the test of equivalence produces sets of scalars providing a unique local geometric characterization, as the spacetime is then characterized by the two discrete sequences $6\geq s_0\geq s_1\geq \ldots \geq s_\pf=s_{\pf+1}\geq 0$ and $0\leq t_0\leq t_1\leq \ldots t_\pf=t_{\pf+1}\leq 4$, the canonical forms used, and the functional relations between the corresponding Cartan invariants up to the final order $\pf+1$.
The ON frame which realizes the constructed canonical form of ${\cal S}_q$, which we call  
a {\em $q$th order invariant frame}, is fixed up to the action of $H_q$. A $\pf$th order invariant frame is just called an {\em invariant frame}, and also a {\em maximally fixed frame}; 
the dimension $s_\pf=\dim(H_\pf)$ indicates continuous residual frame freedom and equals
the dimension of the isotropy group of the spacetime. The dimension of the orbits of the maximal isometry group is $4-t_p$, and the dimension of the group itself is  $4-t_\pf+s_\pf$. From  the perspective of computation, it is of interest to identify the largest value for $\mathfrak{p}$ in the cosmological context; we will discuss this in section \ref{subsec:CKcosm-max}. 

\section{An adaptation to cosmological models}\label{sec:CKcosm}

The Karlhede procedure for local characterization of geometries $(M,{\bf g})$ relies on imposing a canonical form on the curvature tensors to determine a (family of) maximally fixed frame(s), in which the classifying Cartan invariants are computed. In the absence of any  given invariantly-defined directions, this is achieved in the first instance ($q=0$) by using the canonical form of the Weyl or Ricci tensor. 
However, in some cases these tensors cannot be used to maximally fix the frame ($s_0\neq s_\pf$), and higher-order curvature tensors need to be considered. Since canonical forms at higher order may be difficult to find\footnote{The alignment classification for the covariant derivatives of the curvature tensor could be used \cite{ref4, OPP2011}, although this can be difficult to use in practice as well.}, this hinders a fully algorithmic implementation of the method (see section 9.3.2 of \cite{Stephani}). 

Here we study cosmological models, considered as triples $(M,{\bf g},{\bf u})$. We assume that the spacetime geometry $(M,{\bf g})$ is $\mathcal{I}$-non-degenerate, and that the preferred timelike vector field ${\bf u}$  is invariantly-defined from the geometry, in the sense that it can be determined as an eigenvector of some curvature operator, either from the Riemann tensor itself or from a higher order curvature tensor \cite{Hervik:2010rg}. Contrary to the classification of geometries $(M,{\bf g})$ such a field ${\bf u}$ is now additionally given. We can use this extra piece of information to produce a novel and simple Karlhede-like characterization scheme for cosmological models.

In section \ref{subsec:CKcosm-algo} we take a global perspective and describe the essence of the adaptation of the Karlhede scheme to cosmological models. The main task in a Karlhede-like scheme is to define a suitable maximally fixed frame; in our scheme this will be accomplished by using sets consisting only of spatial vectors and PSTF 2-tensors relative to ${\bf u}$, and a general method for normalizing such sets is provided in section \ref{subsec:CKcosm-framefix}. In section  \ref{subsec:algo} we apply this method to the relevant sets to obtain our classification algorithm. The use of Petrov and Segre types and other classifying tools is provided in section \ref{subsec:CKcosm-types}, and we end this section with a discussion of possible maximal values of $\pp$ in section \ref{subsec:CKcosm-max}.  


\subsection{Global perspective and adapted scheme}\label{subsec:CKcosm-algo}

Essentially, the method boils down to replacing the successive covariant derivatives of $R_{abcd}$ in the scheme of section \ref{sec:CKalg} by those of the couple $(u_a,R_{abcd})$. Thus, at order $q$ we {add the $q$th covariant derivative of $u^a$ and} define 
$\Hp_q,\,\sp_q=\dim(\Hp_q)$ and $\tp_q$ for the list
\begin{equation}\label{Sqstar-def}
{\cal S}^*_q=(u_a,R_{abcd}\,;\,\ldots\,;\nabla_{e_q}\cdots\nabla_{e_1}u_a,\,\nabla_{e_q}\cdots\nabla_{e_1}R_{abcd})\,, 
\end{equation}
analogously to $H_q,\,s_q$ and $t_q$ for ${\cal S}_q$. 
In their turn $R_{abcd}$ and $\nabla_b u_a$ can be replaced by the relative Riemann and kinematical quantities \eqref{rel-Ric-mup}-\eqref{rel-Weyl} and \eqref{kin-u} without altering the definitions. Also, it suffices to apply the derivatives \eqref{deriv-time} and \eqref{deriv-proj} instead of the full covariant derivative. As for geometries a canonical form is found at each order $q$, realized by a family of frames determined up to the action of $H^*_q$ (see below), and use can be made of the generalized Ricci and Bianchi identities to find a minimal set of extended classifying invariants (cf.\ section 9.3.2 of \cite{Stephani}). 
As before, we have $\Hp_{q+1}\subset\Hp_q$ (and so $\sp_{q+1}\leq \sp_{q}$) and $\tp_{q+1}\geq \tp_{q}$, and the new procedure stops at order $\pp+1$ where $\pp$ is the smallest number $q$ such that $\sp_{q+1}=\sp_{q}$ and $\tp_{q+1}=\tp_{q}$. 

From a theoretical perspective, our {extended scheme} is a Karlhede-like realization of Cartan's general method (see \cite{Olverbook} and section 9.2 of \cite{Stephani}) applied to triples  $(M,{\bf g},{\bf u})$.
The components of the covariant derivatives of $u^a$, just as those of $R_{abcd}$, can be lifted to functions on the frame bundle; in particular, an expression analogous to (2.87) in \cite{Stephani} holds for the exterior derivatives of the component functions associated to {\bf u}. At order $q$ this gives a lifted map associated to ${\cal S}^*_q$ with rank $\tp_q+6-\sp_q$. We assume the triples to be regular in the sense that both $\tp_q+6-\sp_q$ and $\sp_q$, and thus $\tp_q$, are constant \cite{MilWyl13}. 

Since $u^a$ is invariantly-defined from the geometry, the isotropy group $H_\pf$ of the geometry will leave $u^a$ and all its covariant derivatives invariant, and the components of these tensors in an invariant frame will be functions of Cartan invariants. Hence 
\begin{equation}\label{Hp-tp-equal}
H_\pf=\Hp_{\pp}\,\Rightarrow\,s_\pf=\sp_{\pp},\quad t_\pf=\tp_{\pp}\;. 
\end{equation} 
Since $\sp_q\leq s_q$ and $\tp_q\geq t_q$ for all $q$, this implies that $\pp\leq \pf$. 

The reason why we add the covariant derivatives of $u^a$, and not just $u^a$ itself at zero order, is twofold. 
First, our definition of a cosmological model allows, in principle, for cases where $u^a$ is invariantly-defined from a higher order curvature tensor but 
not from the Riemann tensor itself, which occurs when $\sp_0<s_0$ or $\tp_0>t_0$. For such cases, a method where only $u^a$ would be added at zero order, and not its covariant derivatives, may lead to an invalid classification scheme and stop criterion; in particular, the kinematical quantities, equivalent to $\nabla_b u_a$, then play an essential role, see appendix \ref{app: example} for a hypothetical example with $\sp_0=s_0$ and $\tp_0>t_0$.\footnote{Cosmological models where the Riemann tensor itself, but not its first covariant derivative, is boost-isotropic in some 2D plane would provide examples of $\sp_0<s_0$, since $u^a$ automatically breaks down the boost isotropy. Note that the exterior Schwarschild region is an ${\cal I}$-non-degenerate geometry which has such boost isotropy property; however, it is not the geometry of a cosmological model.} Second, for common cosmological models known and used so far, the fundamental time-like vector field $u^a$ {\em can} be defined from the Riemann tensor, 
and $\sp_q=s_q,\,\tp_q=t_q$ for all $q\geq 0$ (
important examples hereof are perfect fluid models with $u^a$ along the timelike eigendirection of the Ricci tensor; see \eqref{cond-pf}). 
In this case the addition of the covariant derivatives of $u^a$ is not strictly needed. However, kinematical quantities in particular often have more compact expressions than first order curvature tensor components, and as directly available extended invariants in the new scheme (only implicit in the standard one) may be easily incorporated in a minimal list of classifying invariants;
moreover, their use considerably simplifies the classification scheme when $\sp_0>0$ (see proposition \ref{prop:sp=1} and Secs.\ \ref{subsec:algo}
and \ref{sec:examples}).

\subsection{Maximally fixing the frame}\label{subsec:CKcosm-framefix}

For cosmological models $(M,{\bf g},{\bf u})$ we can use the additionally given timelike vector field ${\bf u}$ to find a maximally fixed frame and thus to obtain a complete list of extended invariants in a straightforward and transparent way.

At order $q=0$ of the scheme we dispose of $(u^a,R_{abcd})$. We work with ON frames $(e_0^a,e_i^a)$ {\em which we make {\em {\bf u}-adapted} from the start by setting  $e_0^a=u^a$}. The rotations and reflections in $u^\bot$ constitute the group $\Gu$ of the residual  transformations acting on the so far unspecified ON triad $(e_i^a)$; hence, compared to the algorithm for geometries, the initial vertical dimension of the frame bundle (before looking at the Riemann tensor) is reduced from 6 to 3. The isotropy group at zero order, $\Hp_0$, is a subgroup of $\Gu$, whence so is $\Hp_q$ for any $q$. 
The subgroups of $\Gu$ are well-known and there are three 
possibilities 
at order $q$, depending on the value of $\sp_q$:
\begin{itemize}
	\item $\sp_q=0$. The triad $(e_i^a)$ is fixed at order $q$ up to a finite number of discrete transformations, which constitute $\Hp_q$.  
	\item $\sp_q=1$. At order $q$ only one triad vector, say $e_1^a$, can be fixed (possibly up to reflection) and $\Hp_q$ contains the group of spatial rotations about $e_1^a$. 
	\item $\sp_q=3$. This is the case $\Hp_q=\Gu$ where no direction in $u^\bot$ is preferred at order $q$. 
\end{itemize}
In the last two cases ${\cal S}^*_q$ is LRS, in the last case `maximally LRS'. 
When the final value $\sp_{\pp}=s_\pf$ is 1 or 3 the cosmological model and the underlying spacetime geometry are said to be LRS, resp.\ maximally LRS. 
 
To maximally fix the spatial frame vectors $e_i^a$ in practice we will exploit the relative Riemann quantities (at zero order) and kinematical quantities of {\bf u} (at first order, in case $\sp_0>0$) defined in \eqref{rel-Ric-mup}-\eqref{rel-Weyl} and \eqref{kin-u}. In any {\bf u}-adapted frame the scalars $\mu=T_{00},\,p=\frac 13 T_i{}^i$ and $\Theta=\nabla_i u^i$ are extended invariants. On the other hand, $q^a,\,\dot{u}^a,\omega^a$ are spatial vectors and $\pi_{ab},\,E_{ab},\,H_{ab},\,\s_{ab}$ are PSTF 2-tensors. In an arbitrary ON triad these can be represented by $1\times 3$ row matrices and $3\times 3$ tracefree symmetric matrices, respectively. 
For an arbitrary set of such tensors one has the following

\begin{prop}\label{prop:setS}
Consider a set $\S$ consisting of 
(invariantly-defined) 
spatial vectors and PSTF 2-tensors relative to {\bf u}. Let $\HS$ denote its isotropy group and $\ss=\dim(\HS)$. If $\S$ has a non-zero element then $\HS$ is discrete ($\ss=0$) except when there is a unit vector $n^a$ (uniquely defined up to reflection) such that 
\begin{equation}\label{cond-s=1}
S^a\propto n^a\qquad\textrm{and}\qquad S_{ab}\propto h_{ab}-3n_an_b
\end{equation}
for any vector, resp.\ 2-tensor, in $\S$, in which case $\HS$ contains all rotations about $n^a$ and $\ss=1$. If all elements of $\S$ are zero then $\HS=\Gu\,(\ss=3)$.
\end{prop}

Note the geometric meaning of \eqref{cond-s=1}: all non-zero vectors in ${\cal S}$ are aligned with $n^a$, and all non-zero 2-tensors in ${\cal S}$ are degenerate and have a common eigenplane with unit normal $n^a$.  
In the remainder of this paragraph we show how this proposition can easily be proved and put into practice by choosing an ordering for the set $\S$ (i.e., turning $\S$ into a list). There are several cases:

(1) All elements of $\S$ are zero. 
Then $\S$ does not define any preferred directions in $u^\bot$ and the result is trivial. 
Henceforth we assume that $\S$ contains
non-zero elements, and let $\SS$ be the first non-zero tensor in the list.  

(2) $\SS$ is a PSTF 2-tensor with $\IS^3\neq 6\JS^2$ (recall \eqref{def-IS}-\eqref{def-JS}). Then 
$\SS$ has 
distinct eigenvalues $\l_1\neq\l_2\neq\l_3\neq\l_1$ (see section \ref{sec:prelim})
and we use a $\Gu$-rotation to bring it to the diagonal canonical form 
\begin{equation}\label{S-s=0}
\SS\equiv \diag(\l_1,\l_2,\l_3)\quad\textrm{with e.g.}\quad \l_1>\l_2>\l_3,
\end{equation}
such that $(e_i^a)$ becomes an ON eigentriad of $\SS$. This triad is determined up to the reflections $e_i^a\mapsto -e_i^a$ and so $\ss=0$. 

(3) $\SS$ is a spatial vector or a PSTF 2-tensor with $\IS^3=6\JS^2\neq 0$. In the vector case we can normalize $\SS$ to the preferred unit vector $n^a=S^a/|{\bf S}|$, while in the 2-tensor case we take a unit vector $n^a$ along the eigendirection corresponding to the single eigenvalue $-2\JS/\IS$ of $\SS$. Then \eqref{cond-s=1} holds, and in any ON triad $(e_i^a)$ with $e_1^a=n^a$ 
$\SS$ 
takes the canonical form 
\begin{equation}\label{S-s=1}
\SS\equiv \sigma[1\;0\;0]\qquad\textrm{or}\qquad\SS\equiv \l\, \diag(-2,1,1).
\end{equation}
Now either  {\em all} elements of $\S$ take the form \eqref{S-s=1} in such a triad, or not. 

(3a) If this is the case then $(e_2^a,e_3^a)$ cannot be fixed and $\ss=1$. 

(3b) If this is not the case then we use the residual rotations about $e_1^a=n^a$ to normalize the first tensor in the list that does not take the form \eqref{S-s=1}, $\SS'$, to   
\beq\label{S'-vector}
\SS'\equiv [\s'\;\tau\;0],\;\tau\neq 0
\eeq 
when this is a spatial vector, and to
\begin{equation}\label{S'-2-tensor}
\SS'\equiv 
\begin{bmatrix}
-2\l'&\mu&0\\
\mu&\l'+\rho&\nu\\
0&\nu&\l'-\rho
\end{bmatrix},
\quad 
(b1)\;\;
\mu=\nu=0<\rho\quad\textrm{or}\quad
(b2)\;\;
\mu\neq 0
\end{equation} 
when it is a PSTF 2-tensor and (b1) $e_1^a=n^a$ is an eigenvector or (b2) it is not.\footnote{On defining $N_{ab}=h_{ab}-n_an_b$ the forms \eqref{S'-vector} and \eqref{S'-2-tensor}(b) are obtained by aligning $e_2^a$ with $N^a{}_{b}S'^b$ and $N^{ab}S'_{bc}n^c$, respectively; the form \eqref{S'-2-tensor}(a) is diagonal and thus $(e_i^a)$ is an eigentriad of $\SS'$ in this case.}  
In any case $(e_i^a)$ is fixed up to the reflections $e_i^a\mapsto-e_i^a$ and so $\ss=0$.

The procedure yields a normal form for $\S$, which may depend on the chosen ordering when $\ss=0$ but not when $\ss\neq 0$, 
and a 
family of ON triads realizing this form; exhausting reflection freedom leads to a canonical form for $\S$; i.e., all components of all elements of $\S$ are then extended invariants.

\subsection{Algorithm}\label{subsec:algo}

Returning to our adapted scheme we define, at order $q=0$, 
the set
\beq\label{def-S0}
\Snul=\{q^a,\pi_{ab},\,E_{ab},\,H_{ab}\}
\eeq
of relative Riemann quantities. 
One has $\Hp_0=H_{{\cal S}^*_0}=H_{\Snul}$. 
We now provide our classification algorithm, where we describe how for each value of $\sp_0=\dim(\Hp_0)$ the frame eventually gets maximally fixed if we start applying the above normalization procedure to $\Snul$, referring to the possible cases (1), (2), (3a) and (3b) of section \ref{subsec:CKcosm-framefix} for this set. 
For $\sp_0> 0$ 
we use 
the next result, which may be inferred from the analysis in \cite{GoodeWain} (recall \eqref{def-angle}-\eqref{deriv-proj}):

\begin{prop}\label{prop:sp=1} If a preferred unit vector $n^a$ in $u^\bot$ exists such that \eqref{cond-s=1} holds for each element of $\S=\Snul\cup\Spl$ with\footnote{\label{foot:ndot} By $n^an_a=1$ we have $\dot{n}^an_a=0$ such that $\dot{n}^{\langle a\rangle}\propto n^a\Leftrightarrow\dot{n}^{\langle a\rangle}=0$. Also, $R=\mu-3p+4\Lambda$ by \eqref{Ttensor} and \eqref{T-decomp}, such that $\Dtil^a(\mu-3p)=\Dtil^a R$.}  
	\beq\label{def-S+}	
	\Spl=\{\s_{ab},\omega^a,\dot{u}^a,\Dtil^a(\mu-3p),\Dtil_{\langle a}n_{b\rangle},\dot{n}^{\langle a\rangle}\},
	\eeq
	then $H_p$ contains all rotations about $n^a$ and the spacetime is LRS with $s_\pf=1$. 
\end{prop}

\subsubsection{$\sp_0=0$}\label{subsubsec:CKcosm-s0=0}

This is the union of cases 2 and 3b. Choosing an ordering for $\Snul$ we construct the canonical form of the corresponding list and thus fully fix the frame at zeroth order, 
possibly up to reflections $e_i^a\mapsto -e_i^a$. We have  $\sp_q=0,\,\forall q\geq 0$. In principle there are 24 possible orderings for the four elements of $\Snul$, but it is natural to take either $\pi_{ab},q^a$ or $E_{ab},H_{ab}$ as the first two elements (in some order) to obtain a {\em Ricci-preferred}, resp.~{\em Weyl-preferred}, canonical form and frame relative to {\bf u}. 

Starting with $q=0$ one consecutively computes the frame components of the tensors in ${\cal S}^*_q$, and determines the number $\tp_q$ of functionally independent ones (thereby one applies the derivatives \eqref{deriv-time} and \eqref{deriv-proj} and simplifies by using the generalized Ricci and Bianchi identities). If $\tp_{q}=\tp_{q-1}$ we let $q=\pp+1\geq 1$, exhaust remaining reflection freedom, and stop the algorithm. 

\subsubsection{$\sp_0=1$}\label{subsubsec:CKcosm-s0=1} 

This is case 3a. All elements of $\Snul$ satisfy \eqref{cond-s=1} for an up to reflection unique vector $n^a$ and take the form \eqref{S-s=1} in any ON triad $(e_i^a)$ with $e_1^a=n^a$, so 
\beq
\label{s0=1-Ric}
&q^a\equiv [a\;0\;0],\qquad 
\pi_{ab}\equiv (-2b,b,b),\\
\label{s0=1-Weyl}
&E_{ab}\equiv \diag(-2e,e,e),\qquad H_{ab}\equiv \diag(-2h,h,h)
\eeq
where not all of $a,b,e,h$ vanish. 
Now, we verify whether all elements of $\Spl$ also satisfy \eqref{cond-s=1} (i.e., take the form  \eqref{S-s=1} in such a triad). 
\begin{itemize}
\item[(i)] If yes, then the model is LRS with $\sp_{\pp}=s_\pf=1$ by 
proposition \ref{prop:sp=1}. The frame is maximally fixed at zeroth order,  with $\sp_q=1,\,\forall q\geq 0$ and up to the reflection $e_1^a\mapsto -e_1^a$ when $q^a=0$. One computes the frame components of the ${\cal S}^*_q$-tensors
for $q\geq 0$ until $\tp_{q}=\tp_{q-1}$; then we let $\pp+1=q\geq 1$, exhaust the reflection freedom,  and stop the algorithm.  
\item[(ii)] If no, we turn $\Spl$ into a list, take the first element that does not satisfy \eqref{cond-s=1}, and 
normalize it to \eqref{S'-vector} when it is a vector or to \eqref{S'-2-tensor} when it is a 2-tensor. This fully fixes the frame at first order, possibly up to reflections of the form $e_i^a\mapsto -e_i^a$. Hence $\sp_q=0,\,\forall q\geq 1$. One computes the frame components of the ${\cal S}^*_q$-tensors
for $q\geq 1$ until $\tp_{q}=\tp_{q-1}$; then we let $\pp+1=q\geq 2$, exhaust the reflection freedom, and stop the algorithm.
\end{itemize}
Note that \eqref{s0=1-Ric} is equivalent to the energy-momentum tensor \eqref{T-decomp} taking the form 
\begin{equation}\label{T-s0=1}
T_{ab}\equiv 
\begin{bmatrix}
\mu&a&0&0\\
a&p-2b&0&0\\
0&0&p+b&0\\
0&0&0&p+b
\end{bmatrix}
\end{equation}
in any {\bf u}-adapted ON frame with $e_1^a=n^a$, while \eqref{s0=1-Weyl} indicates that the complex tensor
\begin{equation}\label{Q-def}
Q_{ab}=E_{ab}+iH_{ab}
\end{equation}
used in the Petrov classification~\cite{Stephani} takes the form  
\begin{equation}\label{Q-s0=1}
Q_{ab}\equiv \diag(-2\l,\l,\l),\quad \l=e+ih
\end{equation}
in any ON triad $(e_i^a)$ with $e_1^a=n^a$ (see section \ref{subsec:CKcosm-types} for some further discussion). 

\subsubsection{$\sp_0=3\,(\Hp_0=\Gu)$}\label{subsubsec:CKcosm-s0=3} 

This is case 1 in which all elements of $\Snul$ are zero, which is the limit case $a=b=e=h=0$ of \eqref{s0=1-Ric}-\eqref{s0=1-Weyl}. The twice contracted Bianchi identities reduce to  
\begin{align}
\label{conf-flat-Bianchi-contr}
&\Dtil^ap=-(\mu+p)\dot{u}^a,\qquad \dot{\mu}=-(\mu+p)\theta,\quad \Dtil^a\mu=0.
\end{align}
If $\mu+p$ is zero then the model would be a locally Minkowski or (anti-)de Sitter universe (see \eqref{Riedecomp}, \eqref{Ttensor}, \eqref{T-decomp} and \eqref{C-decomp})  
but then $u^a$ cannot represent a preferred congruence of observers determined from the geometry. 
Hence 
\begin{align}
\label{cond-pf}
&\pi_{ab}=0,\;q^a=0\quad (T_{ab}=\mu u_au_b+ph_{ab}),\quad \mu+p\neq 0,\\
\label{cond-confflat}
&E_{ab}=H_{ab}=0\quad (C_{abcd}=0),
\end{align}
such that the models with $\sp_0=3$ are the conformally flat perfect fluids where $u^a$ is 
the up to reflection unique
unit timelike eigenvector $u_R^a$ of the Ricci tensor. These models were exhaustively determined by Stephani~\cite{Stephani} while their Karlhede classi\-fication was discussed in \cite{Bradley1986,Seixas1992b}. 
By the end of section~\ref{subsec:CKcosm-algo} we have $\sp_q=s_q$ and $\tp_q=t_q$ for any $q$. 
The Bianchi identities imply $\sigma_{ab}=0$ and $\omega^a=0$. This leaves 
two cases:
\begin{enumerate}
	\item $\dot{u}^a=0\Leftrightarrow \sp_1=3$. Here the Ricci equation applied to $u^a$ implies
	\begin{equation}\label{conf-flat-Ricci}
	\dot{\theta}=-\frac13\theta^2-\frac12(\mu+3p)+\Lambda,\quad \Dtil^a\theta=0. 
	\end{equation}
	By \eqref{conf-flat-Bianchi-contr} we have $\tp_0\leq 1$ and there are two subcases \cite{Bradley1986}:
	\begin{itemize}
		\item $\theta=0\Leftrightarrow \mu$ constant. Then $\mu+3p=2\Lambda$, so $p$ is constant as well and $\tp_0=\tp_1=0$. {\em This precisely yields the Einstein static universe model}. 
		\item $\theta\neq 0\Leftrightarrow \mu$ non-constant. Then $\tp_0=\tp_1=1$, and {\em this precisely gives the Robertson-Walker cosmologies}. 
	\end{itemize}
For both subcases it follows that $\pp=0$ 
and the frame is maximally fixed at zeroth order with $\sp_q=3,\,\forall q\geq 0$ and $\Hp_{\pp}=\Gu$: no preferred directions in $u^\bot$ exist at all and the model is maximally LRS.   
	\item $\dot{u}^a\neq 0\Leftrightarrow \sp_1=1$. We take $e_1^a$ to be the unit vector $n^a$ along $\dot{u}^a$. Then \eqref{conf-flat-Bianchi-contr} and the Ricci identity applied to $u^a$ implies that all elements of $\Snul\cup\Spl$ satisfy \eqref{cond-s=1} except possibly the spatial vector $\dot{n}^{\langle a\rangle}$. Hence, by proposition \ref{prop:sp=1} and footnote \ref{foot:ndot} there are two subcases: 
	\begin{itemize}
		\item[(i)] If $\dot{n}^{\langle a\rangle}=0$ the model is LRS with $\sp_{\pp}=1$; i.e., the frame is maximally fixed at first order and $\sp_q=1,\,\forall q\geq 1$. We compute the ${\cal S}^*_q$-tensor components in the maximally fixed frame for $q\geq 1$ until $\tp_{q}=\tp_{q-1}$; then we let $\pp+1=q\geq 2$ and stop the algorithm.
		\item[(ii)] If $\dot{n}^{\langle a\rangle}\neq 0$ then we take $e_2^a$ along it (i.e., normalize it to \eqref{S'-vector} with $\s'=0$). This fully fixes the frame at second order; i.e., $\sp_q=0,\,\forall q\geq 2$. One computes the ${\cal S}^*_q$-tensor components 
		in this frame for $q\geq 2$ until $\tp_{q}=\tp_{q-1}$, in which case we let $\pp+1=q\geq 3$ and stop the algorithm. 
	\end{itemize}
\end{enumerate} 

At the end of the algorithm the set ${\cal S}^*_{\pp+1}$ is in a canonical form, and the frame components of its elements produce a complete set of extended invariants, fully characterizing the cosmological model. 

\subsection{Petrov and Segre types; additional characteristics}\label{subsec:CKcosm-types}


The Petrov and Segre types are discrete characteristics of a spacetime geometry, referring to the algebraic structure of the Weyl and Ricci tensor, respectively. In the Karlhede scheme for geometries one first calculates the Petrov type (\I, \II, \III, \D, \N\ or \O).
This can be done by means of the complex tensor $Q_{ab}=E_{ab}+iH_{ab}$ 
relative to {\em any} unit timelike vector field $u^a$ (cf.\ \eqref{rel-Weyl} and \eqref{Q-def}). This tensor is PSTF relative to ${\bf u}$. The complex Weyl invariants $I$ and $J$, which are commonly expressed in terms of contractions of copies of the Weyl spinor \cite{Stephani}, can be computed in terms of $Q_{ab}$, cf.\ \eqref{def-IS}-\eqref{def-JS}:
\begin{equation}
I=I_Q=Q^a{}_bQ^b{}_a,\qquad J=J_Q=Q^a{}_bQ^b{}_cQ^c{}_a.
\end{equation}
The Petrov types can now be characterized as follows:~\cite{Stephani,WylCosNat19}
\begin{itemize}
	\item Petrov type with $I^3-6J^2\neq 0$: \I\ (algebraically general type);
	\item Petrov types with $I^3=6J^2\neq 0$: \D\ if $(Q_{\;\;c}^{a}-\lambda h_{c}^{a})(Q_{\;\;b}^{c}+2\lambda h_{b}^{c})=0$ with $\l\equiv -J/I$, else \II; 
	\item Petrov types with $I=J=0$:  \O\ if $Q_{ab}=0$, \N\ if $Q_{\;\;c}^{a}Q_{\;\;b}^{c}=0\neq Q_{\;\;b}^{a}$, else \III\ (in which case $Q_{\;\;c}^{a}Q_{\;\;d}^{c}Q_{\;\;b}^{d}=0\neq Q_{\;\;c}^{a}Q_{\;\;b}^{c}$). 
\end{itemize}
Next, the Weyl tensor is put into a canonical form in accord with the Petrov type; see table 4.2 of \cite{Stephani}. In terms of ON frames $(e_0^a,e_i^a)$ realizing this form and for `diagonal' types \I\ and \D, this implies that $e_0^a$ is a (timelike) {\em Weyl principal vector} $u_C^a$; such a vector  is characterized by the fact that the relative electric and magnetic Weyl operators $E^a{}_b$ and $H^a{}_b$ on ${\bf u}_C^\bot$ commute, and then $(e_i^a)$ is a eigentriad for both, which thus diagonalizes $Q_{ab}$; $u_C^a$ is unique up to reflection in the type \I\ case and is any unit timelike vector lying in the plane $\Sigma_C$ spanned by the Weyl principal null directions in the type \D\ case~\cite{Stephani,WylCosNat19}. For Petrov type \I\ $Q^a{}_b$ has distinct eigenvalues, such that the resulting ON frame $(u_C^a,E_i^a)$ is essentially unique (i.e., determined up to reflections) and called the {\em Weyl principal frame}~\cite{Stephani}. For Petrov types \II\ and \III\ the canonical frame is also realized by an essentially unique frame. For 
types \D\ and \N\ and trivial type \O\ (conformally flat case $C_{abcd}=0$) 
there is residual frame freedom, which is now used to simplify the representation matrix of the Ricci tensor; the Segre type of the Ricci tensor is then  calculated as an additional characteristic. One can also reverse the roles of the Weyl and Ricci tensors, looking first for a canonical form of the latter. 

In our scheme for cosmological models the approach  is essentially different. We immediately take the fundamental vector $u^a$ as $e_0^a$, which is not necessarily a vector that is geometrically `aligned' with the Weyl (or Ricci) tensor, and start to apply the normalization procedure to 3D relative Riemann quantities. Hence both 4D Petrov and Segre types are additional characteristics. 

In this context, let us discuss the compatibility of these types with \eqref{T-s0=1}-\eqref{Q-s0=1}; i.e., with 
$\sp_0>0$. This happens precisely when the Riemann tensor is LRS in some 2D plane $\Pi$
and $u^a\in\Pi^\bot$, and requires specific Petrov and Segre types as well as further alignments between the Ricci and Weyl tensors and $u^a$. The case $\sp_0=3$ precisely covers conformally flat perfect fluids (Petrov type \O, Segre type $[(111),1]$) with $u^a=\pm u_R^a$ (see section \ref{subsubsec:CKcosm-s0=3}), where the Riemann tensor is LRS in the full space ${\bf u}^\bot$. When $\sp_0=1$ the plane $\Pi$ is unique; this happens precisely when the following two conditions hold: 
\begin{itemize}
	\item the Petrov type is \O\ (case $\l=0$ in \eqref{Q-s0=1}) or \D\ with $\Pi=\Sigma_C^\bot$ and $u^a\in \Sigma_C$ a Weyl principal vector (case $\l\neq 0$ in \eqref{Q-s0=1}, cf.\ above);
	\item the Segre type is one of\footnote{Segre type $[(11)Z\overline{Z}]$ can be excluded since it even violates the weak energy condition~\cite{Stephani}. We also did not list the `vacuum type' $[(111,1)]$ (corresponding to $R_{ab}\propto g_{ab}$), which in any case should be combined with Petrov type \D\ for $u^a$ to be invariantly defined (see section \ref{subsubsec:CKcosm-s0=3}), but this gives solutions related to and including the classical black hole metrics spacetimes~\cite{Stephani}.}
	\begin{equation}\label{Segre-s0=1}
	 [(11),2],\;[(11)1,1],\;[(11)(1,1)],\quad[(11,2)],\;[1(11,1)],\;[(111),1],
	\end{equation}
	where for the first three types a unique 2D plane $\Pi_R$ is defined and $u^a\in\Pi=\Pi_R$, while for the 
	last three types a unique null vector $l_R^a$, spacelike vector $m_R^a$, \ timelike vector resp. $u_R^a$ is defined and this vector must belong to $\Sigma_C$ when the Petrov type is \D\ (and spans $\Pi$ together with $u^a$). 
\end{itemize}
In all other cases one has $\sp_0=0$.
In particular, our method does not explicitly distinguish between the 
Petrov types \I, \II, \III, \N, and \D\ with $u^a\notin \Sigma_C$, where in the last two cases $u^a$ breaks down the null rotation, respectively, boost and spatial rotation isotropy contained in the Weyl tensor. 

However, the Petrov type may become apparent when the couple $(E_{ab},H_{ab})$ is normalized first. For instance, when $E_{ab}$ is diagonalized and $H_{ab}$ becomes diagonal in the same {\bf u}-adapted ON frame then this means that the Petrov type is \I\ or \D\ and $u^a$ is a Weyl principal vector (cf.\ above); a subcase hereof is that of a {\em purely electric (magnetic)} model where $u^a$ is a Weyl principal vector, characterized by $E_{ab}\neq 0=H_{ab}$ ($H_{ab}\neq 0=E_{ab}$); many purely electric models are known which are of `aligned' Petrov type \D\ ($u_R^a\in \Sigma_C$), such as the orthogonally spatially homogeneous Bianchi I models, and the Szekeres and the LRS Lema\^itre-Tolman-Bondi and Kantowski-Sachs dust-filled universes (see section \ref{subsec:ex-Szekeres}-\ref{subsec:ex-BI}), while purely magnetic models are more elusive (yet see e.g.\ \cite{WylVdB06} for an example and overview). More generally, (Weyl) purely electric or magnetic spacetimes are characterized by the {\em existence} of some unit timelike vector field relative to which $E_{ab}\neq 0=H_{ab}$ or $H_{ab}\neq 0=E_{ab}$, respectively; a simple criterion for this property is that the Weyl invariant $M\equiv I^3/J^2-6$ is real positive or infinite (where $M=0$ corresponds to Petrov type \D\ and the infinite case to Petrov type \I\ with $J=0$) and the Weyl invariant $I$ is real and strictly positive (electric case) or strictly negative (magnetic case)~\cite{mcintosh1994electric}.   

The Weyl-Petrov, Ricci-Segre, and 
 `mixed' types 
(defined by conditions on certain contractions between 
the Weyl and Ricci
tensors) can be additionally computed and may provide useful geometric information.  Segre or Ellis types~\cite{Visser2018} (in terms of discriminants~\cite{CHDG, CH}) and traces of powers or eigenvalues of rank 2 operators constructed at any order may also be worth computing. 


\subsection{ Maximal value of $\pp$.} \label{subsec:CKcosm-max} It is worthwhile to ask 
how many covariant derivatives are at most needed to classify a cosmological model, as this is related to the computational aspects of the algorithm. Hence we wish to determine theoretical maximal values of $\pp$. Clearly, to obtain such a value requires $\sp_q=s_q$ and $\tp_q=t_q$ for all $q\geq 0$, such that $\pp=\pf$; i.e., we can rely on known results for maximal bounds for spacetime geometries where $u^a$ can be invariantly defined from the Riemann tensor~\cite{Stephani}.   
For $\sp_0=3$ (i.e., conformally flat perfect fluid models with $u^a=\pm u_R^a$) one has $\pp\leq 3$~\cite{Bradley1986,Seixas1992b}.   
For $\sp_0=0$ and 1 theoretical bounds are $\pp\leq 4$ and $\pp\leq 5$, respectively, since the integers $\sp_q$ and $\tp_q$ form a descending, resp.\ ascending, sequence 
ending with $\sp_{\pp}\geq 0$ and $\tp_{\pp}\leq 4$, and one could have 
either $\sp_{q+1}=\sp_q-1,\,\tp_{q+1}=\tp_q$ or $\sp_{q+1}=\sp_q,\,\tp_{q+1}=\tp_q+1$ at each step. The theoretical bound can still be lowered for LRS models ($\sp_0=\sp_{\pp}=1$): 
there are at least three independent Killing vector fields, whence $\tp_{\pp}\leq 2$ and $\pp\leq 3$. 
The bound for $\sp_0=1$ non-LRS models can also be lowered for conformally flat spacetimes of Segre type either $[(11,2)]$ ($\pf\leq 4$) or $[1(11,1)]$ or $[(11)(1,1)]$ ($\pf\leq 1$) by exhaustive analysis~\cite{Stephani}. It would be beneficial to bring down the bounds in other cases as well, and to study whether realistic cosmological models actually allow $\pp\geq 3$. 

\section{Examples: Cosmological Models}\label{sec:examples}

In this section we discuss the application of  
our classification algorithm to several exact cosmological models.\footnote{For an exact cosmological model the timelike vector field {\bf u} is given explicitly as part of the solution. Recall that there is the implicit assumption that the geometry is ${\cal I}$-non-degenerate and {\bf u} is an invariantly-defined vector; this will be shown to be true in all examples. Also, the energy-momentum tensor is specified, in the sense that the Ricci-Segre type is given.}  
In all cases we give the 
consecutive values $\sp_q$ and $\tp_q$ and the stop value $\pp$, and in particular $s_\pf=\sp_{\pp}$ and $t_\pf=\tp_{\pp}$, as well as the dimension $r$ of the maximal isometry group. The first example can be considered as an explicit demonstration of how the algorithm works. 

\subsection{Quasi-spherical Szekeres dust model}\label{subsec:ex-Szekeres}
The quasi-spherical Szekeres metric is a solution to Einsteins field equations for a perfect fluid representing dust ($p=0$)~\cite{szekeres1975class}. It provides an anisotropic and non-spatially homogeneous universe 
model with the fundamental observer's (perfect fluid) 4-velocity {\bf u} given by $\partial_t$. 
The line-element is given by
\begin{equation}\label{Szek-metric}
ds^{2}=-dt^{2} 
+\frac{\left(R_{,z}-\frac{\E_{,z} R}{\E}\right)^{2}}{1+2F}dz^{2} 
+ \frac{R^{2}}{\E^2}\left(dx^{2}+dy^{2}\right),
\end{equation}
such that
\begin{equation}
\E(x,y,z)=\frac{S}{2}\left( \left(\frac{x-P}{S}\right)^{2}+\left(\frac{y-Q}{S}\right)^{2}+1\right),
\end{equation}
where $F,S,P,Q$ are functions of $z$ and
$R=R(t,z)$ is a function satisfying 
\begin{equation}\label{Szek-Rt}
\left(R_{,t} \right)^{2}=2F+\frac{2M}{R}, 
\end{equation}
with $M$ a function of $z$. 
We exclude the following subcases:
\begin{itemize}
	\item $R(t,z)=h(t)m(z),\, M(z)=m(z)^3,\, F(z)=F_0 m(z)^2,\,
	(\frac{dh}{dt})^2=2\left(F_0+1/h\right)$,
	which gives conformally flat, spatially homogeneous dust universes with a $G_6$ group of isometries;
	\item $P,\,Q,\,S$ and $M$ are all constant; this gives either the exterior Schwarschild or the Minkowski metric. 
\end{itemize}
Note that if $P$ or $Q$ is constant it can be set to $0$ by a translation of the $x$- or $y$-coordinate, respectively. Likewise, if $S$ is constant it can be set to $1$ by a simultaneous rescaling of the $x$- and $y$-coordinates. 
 
A natural ON frame adapted to the coordinates is
\begin{equation}\label{Szekeres-frame} {\bf e}_{0} ={\partial_t},~~{\bf e}_{1}= \left( \frac{\sqrt{1+2F}}{R_{,z}-\frac{\E_{,z} R}{\E}} \right){\partial_z},~~
{\bf e}_{2}=\frac{\E}{R}{\partial_x},~~{\bf e}_{3}=\frac{\E}{R}{\partial_y}.
\end{equation}
\noindent We work in this frame and take ${\bf u}={\bf e}_0$ as the fundamental unit timelike vector field. At each point this is the up to reflection unique unit timelike eigenvector ${\bf u}_R$ of the perfect fluid Ricci tensor. 
Hence we have $\sp_q=s_q$ and $\tp_q=t_q$ for all $q\geq 0$. Moreover, the Weyl tensor is purely electric with respect to ${\bf u}$ ($H_{ab}=0$) and the eigenvalues for the ${\bf e}_{2}$ and ${\bf e}_{3}$ directions of the electric Weyl tensor are equal ($\l_2=\l_3=-\tfrac 12\l_1\equiv e$), such that the Petrov type is \D\ and ${\bf u}$ is a Weyl principal vector. Hence the elements of ${\cal S}^r_0$ take the form \eqref{s0=1-Ric}-\eqref{s0=1-Weyl} with $a=b=h=0$; the vector ${\bf n}={\bf e}_1$ is invariantly defined at zeroth order and the Riemann tensor is isotropic  
under the rotations in the plane spanned by ${\bf e}_2$ and ${\bf e}_3$.  
The energy density $\mu$ and the eigenvalue $e$, given by
\begin{equation}\label{Szek-mue}
\mu=\frac{(R/\E)^2(R/\E)_{,z}}{2(M/\E^3)_{,z}},\quad
e=\frac{1}{6}\mu-\frac{M}{R^3}, 
\end{equation}
and it is straightforward to show that these Cartan scalars are functionally independent. Hence at the zeroth order level of the algorithm we have
\begin{equation}\label{Szekeres-q=1}
\tp_0=t_0 = 2,\quad \sp_0=s_0 = 1.
\end{equation}
We are in the case of section \ref{subsubsec:CKcosm-s0=1} and 
proceed to the next step (order $q=1$) of the new algorithm, where we consider the set $\Spl$ defined in \eqref{def-S+}. We first calculate the kinematic quantities of {\bf u} in the ON triad $(e_i^a)$ given by \eqref{Szekeres-frame}. The acceleration and rotation vectors vanish: $\dot{u}^a=\omega^a=0$. The expansion scalar equals 
\beq 
\Theta = \frac{\E(2R_{,t} R_{,z}+ R R_{,tz}) - 3 R_{,t} R \E_{,z} }{R (R_{,z} \E - R \E_{,z})} = 2\frac{R_{,t}}{R}+ \frac{(R_{,t}/\E)_{,z}}{(R/\E)_{,z}},
\eeq
while the shear tensor $\sigma_{ab}$ is diagonal, with components
\beq 
-\frac12 \sigma_{11} =  \sigma_{22} = \sigma_{33} = \frac13 \frac{\E (R_{,t} R_{,z} - R R_{,t,z})}{R(R_{,z} \E - R \E_{,z})}=\frac13\left(\frac{R_{,t}}{R}-\frac{(R_{,t}/\E)_{,z}}{(R/\E)_{,z}}\right). 
\eeq
Thus it takes form \eqref{S-s=1} and so satisfies \eqref{cond-s=1} with ${\bf n}={\bf e}_1$; i.e., the shear tensor is isotropic in the ${\bf e}_2$-${\bf e}_3$ plane as well. 
We also find $\dot{n}^{a}=0$ and $\Dtil^a\mu\propto n^a$. 
Hence, we are in case (i) of the algorithm precisely when the last element $N_{\langle ab\rangle}$ of $\Spl$ is also of the form \eqref{S-s=1}.   
The representation matrix of $N_{ab} = \tilde{\nabla}_{b} n_{a}$ in the triad is 
\beq N_{ab} \equiv \left[ \begin{array}{ccc} 0 & 0 & 0 \\ 
	\frac{\ln(\E)_{,zx}}{(R/\E)_{,z}} & \frac{\sqrt{1+2F}}{R} & 0 \\  
	\frac{\ln(\E)_{,zy}}{(R/\E)_{,z}} & 0 & \frac{\sqrt{1+2F}}{R}  \end{array} \right]\;.
\eeq 

It follows that we are in case (i) precisely when $P,\,Q$ and $S$ are all constant (and then $M$ is non-constant to exclude the Schwarzschild metric); in this case we have $\E=\E(x,y)$ and may set $P=Q=0$ and $S=1$ (see above). This gives the well-known {\em Lema\^{i}tre-Tolman-Bondi dust model}~\cite{tolman1934effect}. 
According to proposition \ref{prop:sp=1} it is LRS in the ${\bf e}_2$-${\bf e}_3$ plane. This requires isometry group orbits of dimension $4-t_\pf\geq 2$, such that $t_\pf\leq 2$. But $t_\pf\geq t_0=2$ by \eqref{Szekeres-q=1}, and so 
\begin{equation}\label{Szekeres-q=p}
t_\pf = 2,\quad s_\pf = 1\quad \Rightarrow \quad r=3,\quad \pp=\pf=0. 
\end{equation}
Hence the algorithm stops at first order, as can be verified by direct calculation. All Cartan scalars are functions of $\mu=\mu(t,z)$ and $e=e(t,z)$, see \eqref{Szek-mue}. The models are LRS II in the Stewart-Ellis classification~\cite{StewartEllis} and exhibit spherical, hyperbolic or planar symmetry, having a maximal group $G_3$ of isometries acting on spacelike 2D orbits tangent to the (integrable) ${\bf e}_2$-${\bf e}_3$ distribution.  

Henceforth we assume that $P,\,Q$ and $S$ are not all constant, such that we are in case (ii) of section \ref{subsubsec:CKcosm-s0=1}. We can fully fix the frame at first order by performing a rotation in the ${\bf e}_2$-${\bf e}_3$ plane such that $N_{\langle ab\rangle}$ obtains the normalized form \eqref{S'-2-tensor}, case (b1). There are two subcases. 
\begin{itemize}
	\item $P$ and $Q$ are constant, $S$ is non-constant. In this case we may set $P=Q=0$. One calculates that all zeroth, first and second order extended invariants are functions of $t$, $z$ and $x^2+y^2$, cf.\ \eqref{Szek-mue}, such that 
	\begin{equation}\label{Szekeres-q=p2}
	\pp=\pf=1,\quad t_\pf= 3,~s_\pf = 0\quad \Rightarrow\quad r=1
	\end{equation}
	and so the maximal group of isometries is $G_1$. It follows from the results of \cite{nolan2007shell,georg2017symmetry} that the models are axially symmetric.
	\item $P$ and $Q$ are not both constant. One finds that  
	\begin{equation}\label{Szekeres-q=p3}
	\tp_1 = t_1=4,~~ \sp_1=s_1 = 0. 
	\end{equation}
	where e.g.\ $\Theta$ and $\s_{22}$ provide two more functionally independent, first order extended invariants. \eqref{Szekeres-q=p3} already gives the highest and lowest possible values for $t_q$ and $s_q$, respectively. To complete the algorithm we must continue to the second order next covariant derivative, and we have 
	\begin{equation}
	\pp=\pf=1,\quad t_\pf= 4,~s_\pf = 0,\quad  r=1
	\end{equation}
	This is the generic case without symmetry~\cite{bonnor1977rotating}.
\end{itemize}
In both subcases the algorithm terminates at second order, and the extended invariants up to this order classify the model completely. 

Let us briefly compare the above approach with the standard Cartan-Karlhede algorithm applied to \eqref{Szek-metric}-\eqref{Szek-Rt}. In the latter it is more cumbersome to decide whether the model is LRS (i.e., Lema\^{i}tre-Tolman-Bondi) or not. If not, we can introduce a rotation with a parameter, $\theta$, in the plane spanned by $e_{2}$ and $e_{3}$ to see that there exists a unique $\theta$ such that the component of the covariant derivative of the electric Weyl tensor ${(\nabla E)}_{223}$ vanishes. In the generic case the covariant derivative of the Riemann tensor provides two more functionally independent components through ${(\nabla R)}_{41433}$ ${(\nabla R)}_{14122}$. However, the expressions are more involved than in the new scheme. 

In terms of a minimal set of Cartan invariants, it has been shown that all quasi-spherical Szekeres dust models can be fully characterized by five real-valued scalar curvature invariants arising from a $1\,+\,3$ split~\cite{Sussman}.

\subsection{The Kantowski-Sachs-like model}\label{subsec:ex-KS}

The Kantowski-Sachs-like metric describes a spacetime which is anisotropic. It is the only spatially homogeneous model that does not have a transitive subgroup. The spatial sections have topology $\mathbb{R} \times S^2$ and admit a 4D symmetry group. The metric can be written as: 
\beq ds^2 -dt^2 + a(t)^2 dx^2 + b(t)^2 ( d\theta^2 + \sin^2 \theta d\phi^2). \label{KSmetric} \eeq
\noindent For a function $f=f(t)$ we will write $\dot{f}\equiv df/dt$. For the choice of metric functions, $a(t) = e^{\sqrt{\Lambda} t}$ and $b(t) = \Lambda^{-\frac12}$, we recover the well-known Kantowski-Sachs dust model which descibes an anisotropic homogeneous universe \cite{KS1966}.

We take a frame adapted to the unit normal to the maximal orbit of the Killing vectors: 
\beq {\bf e}_0 = \partial_t,~{\bf e}_1 = \frac{1}{a} \partial_x,~{\bf e_2}= \frac{1}{b} \partial_\theta,~ {\bf e}_3=\frac{1}{\sin \theta b} \partial_\phi. \label{KSframe} \eeq  

\noindent For ${\bf u} = {\bf e}_0$, the expansion scalar, $\Theta$, is zero while the associated vectors for rotation, $\omega^a$, and acceleration, $\dot{u}^a$, both vanish. The only non-vanishing quantity is the shear tensor, $\sigma_{ab}$, which has non-zero components
\beq\label{KS-sigmaab}
\frac12\sigma_{11} = \sigma_{22} = \sigma_{33} = 
\frac{\dot{a} b - \dot{b} a}{3ba}.
\eeq

\noindent Relative to ${\bf u}$ the magnetic Weyl tensor $H_{ab}$ vanishes, while the electric Weyl tensor has non-zero components 
\beq\label{KS-Eab}
\frac12 E_{11} = E_{22} = E_{33} = 
\frac{\ddot{a} b^2 -\ddot{b}ba- \dot{a} \dot{b} b + \dot{b}^2 a + a}{18b^2 a}.
\eeq 
Hence the Weyl tensor is purely electric and of Petrov type {\bf D}, and {\bf u} is a Weyl principal vector. The Einstein tensor has the following non-zero components:
\begin{equation}\label{KS-Gab}
G_{00} = - \frac{ \dot{b}^2a + 2 \dot{a} \dot{b}b + a}{b^2a},~ G_{11} = - \frac{\dot{b}^2+2\ddot{b}b + 1}{b^2},~ G_{22} = G_{33} = -\frac{\ddot{a}b + \dot{a}\dot{b}+ \ddot{b}a}{ba}. 
\end{equation}

\noindent Take a generic model, in the sense that $a(t)$ and $b(t)$ satisfy $G_{11}\neq -G_{00}\neq G_{22}\neq G_{11}$, and that $G_{00},\,G_{11},\,G_{22}$ and $E_{22}$ are not all constant. Then the frame elements ${\bf e}_0={\bf u}$ and ${\bf e}_1$ are distinct eigenvectors of the Einstein tensor. Hence they are geometrically defined,  {\bf u} can be taken as the fundamental timelike vector field, and $\sp_q=s_q$ and $\tp_q=t_q$ for all $q$. Setting ${\bf n} = {\bf e}_1$, we can use proposition \ref{prop:setS} to show that this spacetime is LRS: the only non-zero tensors in the set $S_0^r \cup S_{+}^r$ defined by \eqref{def-S0}-\eqref{def-S+} are $E_{ab},\,\pi_{ab}$ and $\sigma_{ab}$, and by \eqref{KS-sigmaab}-\eqref{KS-Gab} and \eqref{rel-Ric-piq} these are proportional to 
$h_{ab} + 3 n_a n_b$. As the model is spatially homogeneous we will have 
\begin{equation}\label{KS-pp}
t_\pf = t_0=1,\quad s_\pf=s_0 = 1\quad \Rightarrow \quad r=4,\quad \pp=\pf=0. 
\end{equation}
Hence there is a maximal isometry group $G_4$ acting on 3D spacelike orbits, the model is LRS III in the Stewart-Ellis classification~\cite{StewartEllis}, and 
the classification algorithm will conclude at first order. 

\subsection{Bianchi I model}\label{subsec:ex-BI}
In this example we consider the Bianchi I cosmological model. Here the line-element can be written as
\begin{equation}
ds^{2}=-dt^2+a(t)^{2}dx^{2}+b(t)^{2}dy^{2}+c(t)^{2}dz^{2},
\end{equation}
where $a,b,c$ are functions of time. Again we write $\dot{f}=df/dt$ for $f=f(t)$; we work with the unit normal direction to the $G_3$ isometry orbits, ${\bf u} = \partial_t$, and adapt the following frame: \begin{equation}
{\bf e}_{0}={\bf u}=\partial_{t},~ {\bf e}_{1}=a^{-1}\partial_{x},~{\bf e}_{2}=b^{-1}\partial_{y},~{\bf e}_{3}=c^{-1}\partial_{z}. \label{BianchiIframe}
\end{equation}
\noindent As before, {\bf u} has vanishing acceleration, $\dot{u}^a$, and rotation, $\omega^a$. The shear tensor, $\sigma_{ab}$, takes a diagonal form; its entries and the expansion scalar, $\Theta$, are given by 
\beq 
\begin{aligned} 
& \sigma_{11} = \frac13 \frac{\dot{c} a b + \dot{b} ac -2 \dot{a}bc}{abc},\quad\sigma_{22} = \frac13 \frac{\dot{c} a b - 2 \dot{b} ac + \dot{a}bc}{abc}, \\ 
&\sigma_{33} = -\frac13 \frac{2 \dot{c} a b - \dot{b} ac - \dot{a}bc}{abc},\quad\Theta = -\frac{\dot{c} ab + \dot{b} ac + \dot{a}bc}{abc}. 
\end{aligned}   
\eeq

\noindent Relative to {\bf u} the magnetic Weyl tensor vanishes such that again ${\bf u}$ is a Weyl principal vector. 
The Einstein and electric Weyl tensors also take a diagonal form, with components:
\beq\label{as} \begin{aligned}
	& G_{00} = \frac{\dot{b} \dot{c} a + \dot{a} \dot{c} b + \dot{a} \dot{b}c}{abc}, \\
	 G_{11} = -\frac{\ddot{b} c + b \ddot{c} + \dot{b} \dot{c}}{bc},~
	&G_{22} = -\frac{\ddot{a} c + a \ddot{c} + \dot{a} \dot{c}}{ac},~
	G_{33} = -\frac{\ddot{b} a + b \ddot{a} + \dot{b} \dot{a}}{ab};
\end{aligned} \eeq
\noindent and 
\beq\label{at} \begin{aligned} 
	& E_{11} = 	\frac16 \frac{\dot{a} \dot{b} c + \dot{a}\dot{c} b- 2 \dot{b} \dot{c} a + \ddot{c} ab + \ddot{b} ac - 2\ddot{a} bc}{abc}, \\
	& E_{22} = \frac16 \frac{\dot{b} \dot{a} c - 2 \dot{a}\dot{c} b + \dot{b} \dot{c} a + \ddot{c} ab - 2 \ddot{b} ac + \ddot{a} bc}{abc}, \\
	& E_{33} = - \frac16 \frac{2 \dot{a} \dot{b} c - \dot{a}\dot{c} b - \dot{b} \dot{c} a + 2 \ddot{c} ab - \ddot{b} ac - \ddot{a} bc}{abc}.
\end{aligned} \eeq

\noindent If $G_{ii}\neq -G_{00},\,i=1,2,3$ then ${\bf u}$ is the up to reflection unique unit timelike eigenvector of the Einstein tensor and so is a geometrically defined vector at zeroth order, such that  $\sp_q=s_q$ and $\tp_q=t_q$ for all $q$. Moreover, if there are no relations between the functions $a,b$ and $c$ that lead to two components $G_{ii}$ being equal, the frame \eqref{BianchiIframe} is the essentially {\em unique} extension of ${\bf u}$ (up to reflections) which diagonalizes the Einstein tensor, and so the frame is fully fixed by $G_{ab}$ (or, equivalently, $\pi_{ab}$). Also, if the eigenvalues $E_{ii}$ of the electric Weyl tensor are distinct\footnote{This happens when $(\dot{b}c-\dot{c}b)/a,\,(\dot{c}a-\dot{a}c)/b$ and $(\dot{a}b-\dot{b}a)/c$ are all non-constant.} then the Petrov type is \I\ and ${\bf u}$ is the up to reflection unique Weyl principal vector and the frame \eqref{BianchiIframe} is the essentially unique Weyl principal frame. Assuming one of these cases we have $\sp_0=s_0=0$.
The components of the curvature tensors will only depend on the coordinate $t$, and for a realistic model we also assume that some of the zeroth order components \eqref{as}-\eqref{at} explicitly depend on $t$; i.e., $\tp_0=t_0=1$. Hence the the algorithm stops at first order ($\pp=\pf=0$), and 
\begin{equation}\label{BI-pp}
t_\pf =1,\quad s_\pf = 0,\quad  r=3,
\end{equation}
where the maximal isometry group $G_3$ acts transitively on 3D spacelike orbits. 

%


\subsubsection{Kasner Solution}
As an explicit example of a Bianchi I metric, we study the Kasner metric~\cite{kasner1921geometrical}
\begin{equation}
ds^2=-dt^2+t^{2p_1}dx^2+t^{2p_2}dy^2+t^{2p_3}dz^2,
\end{equation}
which is a vacuum solution if the Kasner exponents $p_{1},p_{2}$ and $p_{3}$  satisfy
$$p_1+p_2+p_3=1 \text{ and } p_{1}^2+p_{2}^2+p_{3}^2=1.$$ 
\noindent The choice of these exponents determines how the three orthogonal directions contract and expand over the course of its evolution. We will parametrize the Kasner exponents by~\cite{Stephani}
\begin{equation} \begin{aligned}
p_{1}(\theta)  &=\frac{1}{3}-\frac{1}{3}\cos(\theta)-\frac{1}{\sqrt{3}}\sin(\theta), \\
p_{2}(\theta)&= \frac{1}{3}-\frac{1}{3}\cos(\theta)+\frac{1}{\sqrt{3}}\sin(\theta) \\
p_{3}(\theta)&=\frac{1}{3}+\frac{2}{3}\cos(\theta). \end{aligned}
\end{equation}

\noindent The metric is flat, with the Weyl tensor equal to zero, exactly for $\theta = 0,\frac{2\pi}{3},\frac{4\pi}{3}.$ Two of the Kasner exponents are equal when $\theta = \frac{\pi}{3},\pi,\frac{5\pi}{3}.$ As the Kasner metric is a vacuum spacetime, we cannot use the Einstein tensor to further fix the frame. Relative to ${\bf u} =  \partial_t$ the magnetic Weyl tensor vanishes, and in the ON frame \eqref{BianchiIframe} the electric Weyl tensor is diagonal, with components 
\begin{equation} \begin{aligned} 
E_{11}&=-\frac{-2\sqrt3\cos(\theta)\sin(\theta)+2\cos^{2}(\theta)-\sqrt3\sin(\theta)-\cos(\theta)-1}{9t^{2}}, \\
E_{22}&=-\frac{2\sqrt3\cos(\theta)\sin(\theta)+2\cos^{2}(\theta)+\sqrt3\sin(\theta)-\cos(\theta)-1}{9t^{2}}, \\
E_{33}&=\frac{2(2\cos^{2}(\theta)-\cos(\theta)-1)}{9t^{2}}. \end{aligned} \end{equation}
Hence we see that if $\theta \notin\{ \frac{k\pi}{3}:k\in\mathbb{Z}\}$ the frame is fixed completely by choosing the frame elements to diagonalize the electric Weyl tensor. Using the general observations about the Bianchi I metrics, the classification algorithm is then concluded at first order, with relevant discrete invariants \eqref{BI-pp}.

\subsection{A tilted Bianchi II solution}\label{subsec:BII}

In previous examples the fundamental timelike vector field ${\bf u}$ was 
defined at zeroth order as the fluid 4-velocity; i.e., at each point it was the unique unit timelike eigenvector of the Ricci or Einstein tensor. Also, the magnetic Weyl tensor relative to {\bf u} vanished, such that the spacetime was purely electric and {\bf u} a Weyl principal observer. 
To provide an example of a cosmological solution to Einstein's field equation where a preferred timelike direction exists that is not directly associated (geometrically aligned) with the curvature tensors 
we consider the class of {\it tilted anisotropic perfect fluid cosmologies}. For such solutions, the fluid 4-velocity ${\bf u}_R$ is tilted with respect to the normals to the hypersurfaces of spatial homogeneity; the coordinate system in which the model is constructed (cf.\ the introduction) is adapted to these hypersurfaces and so not comoving with the fluid. We will consider the following tilted Bianchi II tilted cosmology where the anisotropic fluid is accelerating in one direction~\cite{Hewitt:1991hf}:
\beq\label{Hewitt} ds^2 = - dt^2 + t^{-4q+2}dx^2 + t^{6q} dy^2 +t^{2q}\left(dz+ \left(k t^{2q} +2 n x/\gamma\right) dy \right)^2,\eeq
where the occuring parameters are restricted or related by
\footnote{The numerical typo in the expression for $W^2$ in \cite{Hewitt:1991hf}
	has been corrected in \cite{Stephani}.}  
\begin{align}
\label{Hewitt-paras} 
\begin{aligned}
&\frac{10}{7}<\gamma<2,\quad q= \frac{2-\gamma}{2\gamma},\quad k = \frac{4W}{\gamma p},\\
&W^2 = \frac{(2-\gamma)(11\gamma-10) (7 \gamma -10)}{64 (17 \gamma-18)},\quad n^2=\frac{(2-\gamma)(3\gamma-4)(5\gamma-4)}{17\gamma-18},
\end{aligned}
\end{align}
In \cite{Hewitt:1991hf} Hewitt showed that \eqref{Hewitt}-\eqref{Hewitt-paras} gives the unique tilted Bianchi II perfect fluid solution which is transitively self-similar and possesses an orthogonally-transitive $G_2$. The energy density $\mu$ takes a simple expression, and the pressure $p$ is related to $\mu$ by a $\gamma$-law equation of state: 
\begin{equation}\label{gamma-law}
\mu=\frac{2(2-\gamma)}{\gamma^2t^2},\quad p=(\gamma-1)\mu
\end{equation}
(taking $\Lambda=0$). 
One has $I^3\neq 6J^2$ and so the metric is of Petrov type \I. Hence $s_0=0$ and there is an essentially unique Weyl principal vector ${\bf u}_C\neq {\bf u}_R$ (see below) and Weyl principal frame $({\bf u}_C,{\bf E}_i)$. We note that in the 
present implementation of the standard Cartan-Karlhede algorithm this frame would be employed; however, it is difficult to work with in practice 
(see below). Explicitly the quadratic Weyl invariant is given by
$$
I=(2-\gamma)(3\gamma-2)(3\gamma-4)(4-2\gamma-3i\,n)/(3\gamma^4t^4),
$$
such that by the criterion in section \ref{subsec:CKcosm-types} the spacetime is not purely electric nor magnetic, i.e., no observer measuring $H_{ab}=0$ or $E_{ab}=0$ exists. 

For the cosmological model with metric \eqref{Hewitt}-\eqref{Hewitt-paras} the first choice for the fundamental unit timelike vector field {\bf u} 
arises from the  observation that the metric admits a maximal isometry group $G_3$ generated by the three spatial Killing vector fields $\gamma\partial_x -{2ny} \partial_z,\,\partial_y,\,\partial_z$. 
We can determine {\bf u} as the unit normal to the orbits of the $G_3$. 
An orthonormal frame which is aligned both with the orbit of the orthogonally transitively acting $G_2$ and the normal to the orbits of the $G_3$ is given by~\cite{Hewitt:1991hf}: 
\beq {\bf e}_0 = \partial_t,~{\bf e}_1 = t^{2q-1} \partial_x,~{\bf e}_2 = t^{-q} \partial_z,~{\bf e}_3 = t^{-3q} \left[ \partial_y - \left(kt^{2q} + \frac{2n}{\gamma}x \right) \partial_z \right]. \label{BIIframe} \eeq
\noindent
The vector field ${\bf u}={\bf e}_0$ has vanishing acceleration and rotation, $\dot{u}^a = 0$ and $\omega^a = 0$, while the expansion scalar, $\Theta$, and the frame components of the shear tensor, $\sigma_{ab}$, are given  
\beq 
\begin{aligned}\label{gre}
	&\Theta = -\frac{2q+1}{t},\quad\sigma_{12}=\sigma_{13}=0,\\
	&\sigma_{11} = \frac23 \frac{4q-1}{t},~\sigma_{22} = -\frac13 \frac{q-1}{t},~ \sigma_{23} = -\frac{qk}{t},~\sigma_{33} = -\frac13 \frac{7q-1}{t}.
\end{aligned}
\eeq
Relative to the initial frame \eqref{BIIframe} we have the following (possibly) non-zero components of the Ricci tensor, Weyl electric tensor and Weyl magnetic tensor: 
\beq 
& R_{00}  = -\frac{2q(k^2q+7q-3)}{t^2},~~R_{01} = -\frac{2kqn}{\gamma t^2},~~ R_{11}=  -\frac{2(2\gamma^2 q^2 - \gamma^2 q + n^2)}{\gamma^2t^2}, & \label{are}\\
&R_{22}=R_{33}=\frac{(2-\gamma)^2}{\gamma^2t^2},
\label{bre}
\eeq
\beq & E_{11} = \frac23 \frac{k^2\gamma^2  q^2 -3 \gamma^2 q^2 + \gamma^2 q-n^2}{\gamma^2 t^2}, & \\ 
& E_{22} = 	\frac23 \left( \frac{k^2\gamma^2 q^2 + 6 \gamma^2 q^2 - 2 \gamma^2 q + 2n^2}{\gamma^2 t^2}\right),~~E_{23} = -\frac{kq(2q-1)}{t^2}, &\\ & E_{33} = \frac23 \frac{-2k^2\gamma^2 q^2 - 3 \gamma^2 q^2 +  \gamma^2 q-n^2}{\gamma^2 t^2},& \label{ere}\eeq
\beq & H_{11} = - \frac{2qn}{\gamma t^2},~~H_{22} = -\frac{n(q-1)}{\gamma t^2},~~H_{23} = -\frac{2kqn}{\gamma t^2},~~H_{33} = \frac{n(3q-1)}{\gamma t^2}. & 
\label{fre}
\eeq
\noindent 
Referring to the set $\Snul$ in the classification algorithm, these equations 
imply that
(a) the energy flux vector $q^a$ is aligned with $e_1^a$, and $\pi^a{}_{b}$ has two equal eigenvalues and $e_1^a$ spans the single eigendirection, and (b) $e_1^a$ is a common eigendirection of $E^a{}_b$ and $H^a{}_b$. Moreover, both $E^a{}_b$ and $H^a{}_b$ have distinct eigenvalues and cannot be diagonalized simultaneously since ${\bf e}_0\neq {\bf u}_C$ (see below). 
Hence, 
for any ordering for $\Snul$ where $E_{ab}$ comes before $H_{ab}$ the normalization procedure of section \ref{subsec:CKcosm-framefix} leads to the same canonical form for $\Snul$, realized by the essentially unique ON eigenframe of $E^a{}_b$; this frame is obtained by a specific rotation
\beq\label{spin} 
{\bf e}_2' = \cos(S) {\bf e}_2 + \sin(S) {\bf e}_3, \quad {\bf e}_3' = -\sin(S) {\bf e}_2 + \cos(S) {\bf e}_3 
\eeq 
about ${\bf e}_1$, and relative to it $E_{ab}$, but not $H_{ab}$, takes a diagonal form  while $q^a$ and $\pi_{ab}$ take the form \eqref{s0=1-Ric} with $ab\neq 0$. If $H_{ab}$ comes before $E_{ab}$ in an ordering for $\Snul$ this gives the second possible (and similar) canonical form. 

The timelike vector field ${\bf e}_0$ can be determined in a coordinate independent manner since it is normal to the orbits of the $G_3$. While it appears to be unrelated to the curvature tensors  we will show that it is a constant linear combination of vectors that are invariantly defined from the Riemann tensor, and thus invariantly defined itself. 
Here a key fact is that the above conditions (a) and (b) are preserved under boosts 
\beq\label{boost}
{\bf e}_0' = \cosh(\theta) {\bf e}_0 + \sinh(\theta) {\bf e}_1,\quad  {\bf e}_1' =  \sinh (\theta) {\bf e}_0 + \cosh(\theta) {\bf e}_1. 
\eeq
in the ${\bf e}_0$-${\bf e}_1$ plane and rotations \eqref{spin} in the ${\bf e}_2$-${\bf e}_3$ plane.

The second natural choice for {\bf u} is the perfect fluid 4-velocity ${\bf u}_R$. 
It is obtained as the vector ${\bf e}'_0$ for the boost \eqref{boost} with
\begin{equation}\label{tanh}
\tanh(\theta)=\sqrt{\frac{(3\gamma-4)(7\gamma-10)}{(11\gamma-10)(5\gamma-4)}},
\end{equation}
where $\theta$ is the angle of tilt between ${\bf u}_R$ and the previous choice ${\bf e}_0$~\cite{Hewitt:1991hf}. Since $\theta\neq 0$ we have ${\bf e}_0\neq {\bf u}_R$. 
Relative to ${\bf e}'_0={\bf u}_R$ we now have $q'^a=\pi'_{ab}=0$ and find back \eqref{gamma-law}, but as before the respective operators $E'^a{}_b$ or $H'^a{}_b$ relative to ${\bf u}_R$ have distinct eigenvalues and cannot be diagonalized simultaneously since ${\bf u}_R\neq {\bf u}_C$. 
Again this leads to two possible canonical forms for ${\cal S}_0'^{r}$ and corresponding uniquely defined frames, and since condition (b) is preserved this implies that {\em ${\bf e}'_1$ defined by \eqref{boost}-\eqref{tanh} is characterized as the up to reflection unique unit common eigenvector of the electric and magnetic Weyl operators $E'^a{}_b$ and $H'^a{}_b$ relative to ${\bf u}_R$.}\footnote{As an alternative characterization, ${\bf e}_1'$ is aligned with the acceleration vector of the fluid.}
Hence, by inverting \eqref{boost}-\eqref{tanh} we see that the first choice ${\bf e}_0$ is indeed a constant linear combination of the vectors ${\bf e}'_0$ and ${\bf e}'_1$ which are invariantly defined from the Riemann tensor, as claimed.


A third possible choice for {\bf u} is the Weyl principal vector ${\bf u}_C$. This is the unique unit vector for which the relative electric and magnetic Weyl tensors can be simultaneously diagonalized, yielding the essentially unique Weyl principal frame $({\bf u}_C,{\bf E}_i)$. This frame is found by performing a spin-boost \eqref{spin}-\eqref{boost} with parameters $\theta$ and $S$ which solve the complex-valued equation
\beq\label{boost-uC} e^{4(\theta+iS)}=\frac{(r_-+2i-i\gamma)(r_--6i+5i\gamma)}{(r_++2i-i\gamma)(r_+-6i+5i\gamma)}\quad\textrm{where}\quad r_{\pm}=2(n\pm4W).
\eeq
Obviously $\theta\neq 0$, and the value defined by \eqref{tanh} does not satisfy \eqref{boost-uC}; 
hence ${\bf e}_0\neq {\bf u}_C\neq {\bf u}_R$ and so the Einstein operator $G^a{}_b$ does not diagonalize in $({\bf u}_C,{\bf E}_i)$. However, since conditions (a) and (b) are preserved under spin-boost, the quantities $q'^a$ and $\pi'_{ab}$ relative to ${\bf e}'_0={\bf u}_C$ still have the form \eqref{s0=1-Ric}; but one has $a\neq 0$ such that $0\neq q'^a\propto {\bf e}'_1$ and so ${\bf u}_C$ is indeed not the timelike eigenvector ${\bf u}_R$ of $G^a{}_b$. Note that this singles out ${\bf e}'_1$ and thus, on inverting \eqref{boost} with \eqref{boost-uC}, again proves that ${\bf e}_0$ is a constant linear combination of vectors invariantly-defined from the Riemann tensor. It also follows that the choice ${\bf u}={\bf u}_C$ leads to a single canonical form for the relative set ${\cal S}_0'^r$ (realized only in $({\bf u}_C,{\bf E}_i)$), whatever ordering we take for its elements at the start of the normalization procedure in section \ref{subsec:CKcosm-framefix}. It can be shown that the relative operators $E'^a_{~b}$ and $H'^a_{~b}$ both have distinct eigenvalues; however, solving \eqref{boost-uC} is involved and leads to intricate expressions for these eigenvalues and other extended invariants. This makes the choice ${\bf u}={\bf u}_C$ unfavorable, apart from being physically less clear.  
  
In each of the three approaches a fully fixed frame is obtained and so $\sp_0=0$, as we knew a priori since $s_0=0$. 
Clearly, the number of functionally independent invariants at zeroth order is $\tp_0 = t_0 = 1$. By computing the covariant derivative of the zeroth order tensors we find that no new functionally independent invariants appear. Of course this is a priori the case since the solution admits a $G_3$ and so:
$$ \sp_1 = s_1 = 0,~~ \tp_1 = t_1 = 1.$$
Thus, the algorithm concludes at the first iteration ($\pp=\pf=0$) and the relevant discrete invariants are: 
$$  t_\pf = 1,~~s_\pf=0,~~ r=3.$$ 


\subsection{A $G_2$ Solution}\label{subsec:ex-G2}

As a last example, we will consider a triple $(M, {\bf g}, {\bf u})$ related to the class of non-diagonal separable $G_2$ spacetimes found in \cite{SM} with metric given by
\begin{equation}\label{class of metrics}
ds^2:= T_{f}^2F^2(-dt^2+dx^2) + T_{g}G\big[T_{p}Pdy^2+\frac{1}{T_{p}P}(dz+T_{w}Wdy)^2\big],
\end{equation}
where $T_{f},T_{g},T_{p}$ and $T_{w}$ are non-zero functions of $t$ (for which we write $\dot{f}=df/dt$) and $F,G,P$ and $W$ are non-zero functions of $x$ (for which we write $g'=dg/dx$).

A convenient ON co-frame for the general class of metrics is given by
\begin{equation}\begin{aligned} 
& {\bf w}^{0}=T_{f}Fdt,\quad {\bf w}^{1} = T_{f}Fdx, \\  
& {\bf w}^{2}=\sqrt{T_{g}T_{p}}\sqrt{GP}dy, \quad {\bf w}^{3}=\sqrt{\frac{T_{g}G}{T_{p}P}}(dz+T_{w}Wdy). \end{aligned} \label{G2-frame}
\end{equation}

\noindent {\em Kinematical frames.}
Previously, we started by putting the relative Riemann quantities into some canonical form, in accord with the global perspective of the new algorithm (see section \ref{subsec:CKcosm-algo}). It is sometimes advantageous to instead normalize the kinematical quantities $\s_{ab},\,\omega^a,\,\dot{u}^a$ first, in some order, by the general procedure described after proposition \ref{prop:setS}. This comes down to switching $\nabla_b u_a$ and $R_{abcd}$, and gives a (family of) {\em kinematical frames}. We shall use such a choice of frame in the current example and note that this affects the value of $\sp_0$ but not of $\sp_1$. 
\vspace{ 2 mm}

Let $({\bf e}_0,{\bf f}_1,{\bf f}_2,{\bf f}_3)$ be the dual ON vector frame. We will choose the preferred unit timelike vector field to be ${\bf u} = {\bf e}_0=(T_f F)^{-1} \partial_t$, 
which is associated with  separation of variables in comoving coordinates. This vector field is not necessarily algebraically defined from the Riemann tensor or its covariant derivatives, but could in principle be invariantly defined using the Cartan-Karlhede algorithm in terms of extended Cartan invariants and an associated invariant frame. However, we will show that under two simple `genericity' conditions 
the frame can be fully fixed by using the kinematical quantitites of {\bf u}.

The rotation vector, $\omega^a$, is identically zero, while the acceleration vector is 
\begin{equation}\label{G2-acc}
\dot{u}^a:=u^{b}\nabla_{b}u^{a}=\frac{1}{FT_{f}}\frac{F^{\prime}}{F} f_{1}^{a}.
\end{equation}
and relative to \eqref{G2-frame} 
the non-zero components of the shear tensor are, respectively:
\begin{equation}\label{G2-shear}\begin{aligned}
&\sigma_{11} =\frac{1}{FT_{f}}\left(\frac{2}{3}\frac{\dot{T}_{f}}{T_f}-\frac{1}{3}\frac{\dot{T}_{g}}{T_g}\right), \quad \sigma_{22}:=\frac{1}{FT_{f}}\left(-\frac{1}{3}\frac{\dot{T}_{f}}{T_{f}}+\frac{1}{6}\frac{\dot{T}_{g}}{T_g}+\frac{1}{2}\frac{\dot{T}_{p}}{T_p}\right), \\
&  \sigma_{33}=\frac{1}{FT_{f}}\left(-\frac{1}{3}\frac{\dot{T}_{f}}{T_{f}}+\frac{1}{6}\frac{\dot{T}_{g}}{T_{g}}-\frac{1}{2}\frac{\dot{T}_{p}}{T_{p}}\right), \quad \sigma_{23}=\frac{1}{2}\frac{1}{FT_{f}}\frac{W}{P}\frac{\dot{T}_{w}}{T_{p}}.
\end{aligned}     
\end{equation}
We assume that $F'$ vanishes nowhere in the considered neighbourhood of space-time. 
Then, by \eqref{G2-acc}, the acceleration vector is non-zero and aligned with $f_1^a$. Hence, if we order ${\cal S}$ to the list $[\omega^a,\dot{u}^a,\sigma_{ab}]$ then $S^a=\dot{u}^a$ is the first non-zero element, and we take ${\bf n}={\bf f}_1$ as the new ON frame vector ${\bf e}_1$. The shear tensor $S'_{ab}=\sigma_{ab}$ is the final element of the list. 
From \eqref{G2-shear} we see that $e_1^a$ is an eigenvector of $\sigma^a{}_{b}$ acting on ${\bf u}^\perp$. Hence $\sigma^a{}_{b}$ acts as a self-adjoint operator on the subspace $\{{\bf u}, {\bf e_1}\}^{\perp}$, with representation matrix 
\begin{equation*}
\begin{bmatrix}
\sigma_{22} & \sigma_{23}\\
\sigma_{23} & \sigma_{33}
\end{bmatrix}
\end{equation*} 
in the basis $({\bf f}_2,{\bf f}_3)$. The eigenvalues of this matrix are given by
\begin{equation*}
\alpha_{\pm}=\frac{\sigma_{22}+\sigma_{33}\pm\sqrt{(\sigma_{22}-\sigma_{33})^2+4\sigma_{23}^2}}{2}=\lambda'\pm\rho.
\end{equation*}
Here $\lambda'$ is the trace of the operator and $\rho$ the discriminant of its characteristic function, the symbols referring to \eqref{S'-2-tensor}. Hence, if 
\begin{equation}\label{G2-cond-gen}
4\rho^2=(\sigma_{22}-\sigma_{33})^2+4\sigma_{23}^2=\left(\frac{1}{FT_{f}}\frac{\dot{T}_{p}}{T_p}\right)^2+\left(\frac{1}{FT_{f}}\frac{W}{P}\frac{\dot{T}_{w}}{T_{p}}\right)^2
\end{equation}
is zero we are in case (3a) of the normalization procedure, where the operator is a multiple of the identity and thus the frame cannot be fully fixed. By \eqref{G2-cond-gen} and $W\neq 0$ this is the case if and only if $\dot{T}_{p}=\dot{T}_{w}=0$. However, if we assume that ${T}_{p}$ and ${T}_{w}$  vanish nowhere simultaneously we have $\rho\neq 0$ and are in case (3b1) at each point; the procedure now yields the ON eigenbasis $({\bf e}_2,{\bf e}_3)$ of the operator, which is unique up to the reflections ${\bf e}_2\mapsto -{\bf e_2}$ and ${\bf e}_3\mapsto -{\bf e_3}$. Explicitly, we have 
\begin{equation*}
{\bf e}_2=\cos(\theta){\bf f}_{2}+\sin(\theta){\bf f}_{3},\quad {\bf e}_3 =-\sin(\theta){\bf f}_2 + \cos(\theta){\bf f}_{3}
\end{equation*}
where $\theta=\theta(t,x)$ is given by 
\begin{equation*}
\theta=0\;\;\textrm{if}\;\;\sigma_{23}=0\qquad\textrm{and}\qquad \tan\theta = \frac{\alpha_{+}-\sigma_{22}}{\sigma_{23}}\;\;\textrm{if}\;\;\sigma_{23}\neq 0.
\end{equation*}

\noindent {\bf Summary :}	{\it Under two genericity conditions, namely (a) $F'$ does not vanish and (b) $\dot{T}_p$ and $\dot{T}_w$ do not vanish simultaneously, the frame completion of {\bf u} can be fully fixed at each point by the acceleration vector and shear tensor of {\bf u}, up to two reflections. }

Inspection of $R_{abcd}$ in the coframe dual to $({\bf e}_0,{\bf e}_1,{\bf e}_2,{\bf e}_3)$ 
gives, e.g.,
\begin{equation}
\tensor{R}{_0_1_1_0} =\frac{F^{\prime\prime}F{T_{f}}^2-F^2\ddot{T}_{f}T_f+F^2{\dot{T}_{f}}^2-F^{\prime 2}{T_{f}}^2}{{T_{f}}^4F^4},
\end{equation}
\begin{equation}
\tensor{R}{_0_1_2_3}=\frac{1}{2}\frac{T_{g}G(P^{\prime}T_{p}W\dot{T}_{w}-{\bf w}^{\prime}T_{w}\dot{T}_{p}P)}{{T_{f}}^2F^2{T_{p}}^2P^{2}T_{g}G}.
\end{equation}
These components are generically functionally independent. Since the component functions of the metric are functions of $x$ and $t$ alone, the maximal number of functionally independent components of the curvature tensor is two.  

While the frame was fixed using first order quantities in the modified algorithm, in general the dimension of the isotropy group at zeroth order is zero; i.e., $s_0^* = s_0 = 0$ and the number of functionally independent invariants at zero order is: $t^*_0 = t_0 = 2$. By the above these are already the lowest, resp.\ highest, possible values. Hence, when the algorithm continues to the next iteration, 
no new functionally independent invariants appear, and so 
$$s^*_1 = s_1 = 0,\quad t^*_1 = t_1 = 2.$$
Thus, the algorithm concludes at the first iteration ($\pp=\pf=0$) and the relevant discrete invariants are: 
$$
t_\pf=2,\quad s_\pf=0,\quad r=2.
$$

As we have noted above, the timelike direction ${\bf u}$ may not be geometrically defined explicitly from the Riemann tensor or its covariant derivatives. However, the representation of the Einstein tensor in the frame $({\bf e}_0,{\bf f}_i)$ or $({\bf e}_0,{\bf e}_i)$ has a 2+2 block structure; hence, depending on the concrete model 
it is possible to choose a ${\bf u}$ that will be an eigenvector of the stress-energy tensor by either applying a boost in the ${\bf e}_{0}$-${\bf e}_{1}$ plane or by requiring that ${\bf u} = {\bf e}_{0}$ is the velocity for a {\em perfect fluid} field. For example, in \cite{SM} a subclass of non-diagonal separable $G_2$ perfect fluid solutions to Einsteins field equations were found by requiring that the initial frame \eqref{G2-frame} consists of eigenvectors of the Einstein tensor for a perfect fluid solution and solving the resulting differential equations. In this subclass of metrics, the frame \eqref{G2-frame} will be uniquely defined by the Ricci tensor.

\section{Summary and discussion}

In this paper we have outlined a Cartan-Karlhede-like algorithm for cosmological models. Such a model is formally defined as a triple $(M, {\bf g}, {\bf u})$ in terms of a particular $\mathcal{I}$-non-degenerate spacetime, $(M, {\bf g})$ and a preferred unit timelike vector field, ${\bf u}$. This is due to the requirement that a unique locally defined timelike congruence exists in a cosmological model which represents a family of fundamental observers. Typically, this congruence is associated with the 4-velocity of the averaged matter in the model; however, other choices for the timelike vector fields may arise from the geometric structure of the spacetime. 


In any cosmological model, the additional preferred structure {\bf u} yields a covariant $1\,+\,3$ split of spacetime. The proposed algorithm 
works with orthonormal frames having {\bf u} as the timelike frame vector. Hence the group of permissible frame transformations is reduced from the six-dimensional Lorentz group to the three-dimensional compact group of spatial rotations in the rest space of {\bf u}. 
The rotations are used to align the spatial frame vectors to two simple types of invariantly-defined spatial tensors,
namely spatial vectors and spatial tracefree symmetric 2-tensors. By treating the covariant derivatives of ${\bf u}$ on the same footing as the curvature tensors, we incorporate the kinematical quantities in particular. Doing so we have shown that at most ten tensors of these two types (namely the elements of $\Snul$ and $\Spl$) are needed to maximally fix the orthonormal frame. This makes the algorithm very easy to implement. We can now construct a list of extended invariants relative to the fundamental observers with normalized 4-velocity {\bf u}, simply by considering scalars and components of tensors relative to a maximally fixed frame. This list of invariants will characterize the model completely in a coordinate independent manner, and provide insight into the physical description of the cosmological models.

We emphasize that the choice of ${\bf u}$ is not arbitrary. The algorithm will only provide permissible invariants for a cosmological model when the preferred timelike vector field is employed. As the spacetimes involved in cosmological models are $\mathcal{I}$-non-degenerate, {\bf u} can in principle be determined  as an eigenvector of some curvature operator, either from the Riemann tensor itself or from higher derivatives~\cite{Hervik:2010rg}. If the timelike vector field is defined as some operator constructed from the Riemann tensor then the discrete invariants recording the number of functionally independent invariants and dimension of the isotropy group at each iteration of the algorithm match  those of the Cartan-Karlhede algorithm, i.e., $s^*_{q^*} = s_q$ and $t^*_{q^*} = t_q$. 
If ${\bf u}$ arises from an operator involving higher order curvature tensors then  $s^*_{q^*} \leq s_q$ and $t^*_{q^*} \geq t_q$ for all $q$ which implies that $\mathfrak{p}^* \leq \mathfrak{p}$ so that the modified algorithm could potentially conclude before the Cartan-Karlhede algorithm.

There are several avenues for future work on cosmological triples. In four dimensions the application of the resulting invariants to the  physical interpretation of  cosmological models should be explored, for example by algebraically classifying the Einstein tensor using observer based invariants. The concept of a cosmological model can be extended to higher dimensions, and a higher-dimensional algorithm relying on an invariantly defined unit timelike vector field can be used to classify brane cosmologies \cite{Harko:2002hv, Quevedo:2002xw}. The concept of a general triple $(M, {\bf g}, {\bf u})$, where $(M,{\bf g})$ could be $\mathcal{I}$-degenerate and the unit timelike vector field ${\bf u}$ is imposed from some other mathematical considerations is more appropriate for studying higher dimensional cosmological models. However, the concept of a general triple should be investigated further as their classification will differ from triples representing cosmological models. As an example, for a general triple in four dimensions, \eqref{Hp-tp-equal} may no longer be valid and the stopping conditions for a corresponding modified algorithm will differ 
from those of the Cartan-Karlhede algorithm. 

\appendix

\section{Appendix A: The importance of computing kinematical quantities}\label{app: example}

Referring to section \ref{subsec:CKcosm-algo} we show that a method where only $u^a$ is added at zeroth order, and not its covariant derivatives at higher orders, may sometimes lead to an incorrect classification scheme. We do this by constructing a class of hypothetical models where, if the kinematical quantities (first covariant derivative of $u^a$) were not computed, such a method would lead to a false-stop. 

We first restrict the spacetime geometry $(M,{\bf g})$ of the models. We assume they have $s_0=0$, and hence $s_q=0$ for any $q$, and a $t_q$-sequence of the form
$$
(t_0,t_1,t_2,\ldots)\;=\;(0,1,m,\ldots)\quad \textrm{where}\quad m\geq 2. 
$$
The condition $s_0=0$ means that an ON invariant tetrad $(E_\alpha^a)=(E_0^a,E_i^a)$ can be uniquely defined from the Riemann tensor, up to discrete transformations; for instance, if the Petrov type is \I, \II\ or \III , the Weyl principal tetrad can be taken (see section \ref{subsec:CKcosm-types}). We emphasize that $E_0^a\neq \pm u^a$ (see below for the assumption on $u^a$). Next, $t_0=0$ tells that, in such an invariant tetrad, the Riemann tensor components are all constant; i.e., the geometry is {\em curvature homogeneous of order 0}~\cite{MilWyl13}.
Hence the components of the first covariant derivative of the Riemann tensor in the tetrad can be written as  
\begin{equation}\label{ex-Rie1}
\nabla_{\epsilon}R_{\alpha\beta\gamma\delta}
=2\Gamma^{\zeta}{}_{[\alpha|\epsilon}R_{\beta]\zeta\gamma\delta}+2\Gamma^{\zeta}{}_{[\gamma|\epsilon}R_{\alpha\beta|\delta]\zeta}
\end{equation}
where $\Gamma^{\alpha}{}_{\beta\gamma}=(E_\gamma^a\nabla_{a} E_\beta)^\alpha$ are the connection coefficients of the tetrad. Consider \eqref{ex-Rie1} as a linear system in the connection coefficients. Given $s_0=0$, the argument given after Eq.\ (34) of \cite{MilPel09} implies that all connection coefficients appear explicitly, and that the system can be solved to obtain expressions for all connection coefficients, linear in the components $\nabla_{\epsilon}R_{\alpha\beta\gamma\delta}$ and rational in the constants $R_{\alpha\beta\gamma\delta}$. The condition $t_1=1$ implies that all components $\nabla_{\epsilon}R_{\alpha\beta\gamma\delta}$ depend on only one Cartan invariant, say $x$, and hence so do all connection coefficients: $\Gamma^{\alpha}{}_{\beta\gamma}=\Gamma^{\alpha}{}_{\beta\gamma}(x)$. Now we look at the components of the second covariant derivative of the Riemann tensor: 
\begin{equation}\label{ex-Rie2}
\begin{aligned}
\nabla_{\chi}\nabla_{\epsilon}R_{\alpha\beta\gamma\delta}
=&\; 
(\nabla_{\epsilon}R_{\alpha\beta\gamma\delta})'\,E_\chi(x)\\
&-\Gamma^{\zeta}{}_{\epsilon\chi}\nabla_{\zeta} R_{\alpha\beta\gamma\delta} 
+2\Gamma^{\zeta}{}_{[\alpha|\chi}\nabla_\epsilon R_{|\beta]\zeta\gamma\delta}+2\Gamma^{\zeta}{}_{[\gamma|\chi}\nabla_{\epsilon}R_{\alpha\beta|\delta]\zeta},
\end{aligned}
\end{equation}
where we write $f'$ for the derivative of a function $f$ of $x$ and $E_\chi(x)$ for the action of the directional derivative $E_\chi$ on the scalar function $x$. The derivatives $(\nabla_{\epsilon}R_{\alpha\beta\gamma\delta})'$ and the terms on the second line of \eqref{ex-Rie2} only depend on $x$. The condition $t_2\leq2$ now tells that there is at least one component $\nabla_{\tilde{\chi}}\nabla_{\tilde{\epsilon}}R_{\tilde{\alpha}\tilde{\beta}\tilde{\gamma}\tilde{\delta}}$ which is not a function of $x$ only. Hence, 
we have that $E_{\tilde{\chi}}(x)$ is not a function of $x$ only.

Consider now a hypothetical model $(M,{\bf g},{\bf u})$ with the following properties: (a) the geometry $(M,{\bf g})$ is as above; (b) the vector field $u^a$ of fundamental observers can be defined from the first order covariant derivative of the Riemann tensor, and not from the Riemann tensor itself, in the sense that the expansion of $u^a$ in the basis $(E_\alpha^a)$ has coefficients $f^a$ which only depend on $x$, 
\begin{equation}\label{ex-u-exp}
u^a=f^\alpha\,E_\alpha^a,\quad f^\alpha=f^\alpha(x),
\end{equation}
and where $f^{\alpha}$ is non-constant ($(f^\alpha)'\neq 0$) for at least one $\alpha=\bar{\alpha}$. 
Suppose now that we had developed a classification procedure which only adds $u^a$ at order zero; i.e., we would build extended invariants from the lists
\begin{equation*}
{\cal S}^{\#}_0=(u^a,R_{abcd}),\quad {\cal S}^{\#}_q=(u_a,R_{abcd}\,;\nabla_e R_{abcd}\,\ldots\,;,\,\nabla_{e_q}\cdots\nabla_{e_1}R_{abcd}),\,\forall q\geq 1.
\end{equation*}
Evidently the isotropy group of ${\cal S}^{\#}_q$ has dimension $s^{\#}_q$ for any $q\geq 0$. On writing $t^{\#}_q$ for the number of functionally independent components in any canonical form for ${\cal S}^{\#}_q$ we would have 
$$t_0^{\#} = t^{\#}_1 = 1.$$
Hence, the procedure would erroneously stop at $\pf'+1 =1$. To avoid such a false-stop, general Cartan theory predicts that it is safe to consider our general procedure based on the lists ${\cal S}^*_q$ instead. Indeed, with \eqref{ex-u-exp} we have, for the above values $\tilde{\chi}$ and $\bar{\alpha}$, that
$$
\nabla_{\tilde{\chi}} u^{\bar{\alpha}} = f^{\bar{\alpha}} E_{\tilde{\chi}}(x)+\Gamma^{\bar\alpha}{}_{\beta\tilde{\chi}}f^\beta
$$
is not a function of $x$ only, such that 
$$
\tp_0=1,\quad \tp_2\geq 2
$$
and our algorithm continues. This is essentially due to the fact that $\nabla_b u_a$ takes a relevant part of $\nabla_{f}\nabla_e R_{abcd}$ into account when $u^a$ is defined from $\nabla_e R_{abcd}$. 

\section*{Acknowledgements} 

We would like to thank Sigbj\o rn Hervik for useful discussions during this project. The work was supported by NSERC of Canada (A.C.), and through the Research Council of Norway, Toppforsk grant no. 250367: Pseudo- Riemannian Geometry and Polynomial Curvature Invariants: Classification, Characterisation and Applications (L.W., M.A. and D.M.).

\bibliographystyle{unsrt-phys}
\bibliography{cosmocartanRefs1}

\begin{thebibliography}{10}

\bibitem{ellis1999}
G.~F.~R. Ellis and H.~Van~Elst.
\newblock Cosmological models.
\newblock In {\em Theoretical and Observational Cosmology}, pages 1--116.
  Springer, 1999.

\bibitem{ColHerPel09}
A.~Coley, S.~Hervik, and N.~Pelavas.
\newblock {Spacetimes characterized by their scalar curvature invariants}.
\newblock {\em Classical and Quantum Gravity}, 26:025013, 2009.
\newblock \href{http://arxiv.org/abs/gr-qc/0901.0791}{{arXiv:0901.0791
  [gr-qc]}}.

\bibitem{Kundt}
A.~Coley, S.~Hervik, G.~Papadopoulos, and N.~Pelavas.
\newblock Kundt spacetimes.
\newblock {\em Classical and Quantum Gravity}, 26(10):105016, 2009.

\bibitem{siklos1991stability}
S.~T.~C. Siklos.
\newblock Stability of spatially homogeneous plane wave spacetimes. i.
\newblock {\em Classical and Quantum Gravity}, 8(8):1587, 1991.

\bibitem{Hervik:2004qv}
S.~Hervik, R.~van~den Hoogen, and A.~Coley.
\newblock {Future asymptotic behaviour of tilted Bianchi models of type IV and
  VII(h)}.
\newblock {\em Classical and Quantum Gravity}, 22:607--634, 2005.
\newblock \href{http://arxiv.org/abs/gr-qc/0409106}{{arXiv:gr-qc/0409106
  [gr-qc]}}.

\bibitem{CM1991}
J.~Carminati and R.~G. McLenaghan.
\newblock Algebraic invariants of the riemann tensor in a four-dimensional
  lorentzian space.
\newblock {\em Journal of Mathematical Physics}, 32:3135, 1991.

\bibitem{Stephani}
H.~Stephani, D.~Kramer, M.~A.~H. MacCallum, C.~A. Hoenselaers, and E.~Herlt.
\newblock {\em Exact solutions of {Einstein's} field equations, 2nd edition}.
\newblock Cambridge University Press, Cambridge, 2003.
\newblock Corrected Paperback edition, 2009.

\bibitem{Maa1997}
R.~Maartens.
\newblock {Linearization instability of gravity waves?}
\newblock {\em Physical Review D}, 55:463--467, 1997.
\newblock \href{http://arxiv.org/abs/astro-ph/9609198}{{arXiv:9609198
  [astro-ph]}}.

\bibitem{HerOrtWyl13}
S.~Hervik, M.~Ortaggio, and L.~Wylleman.
\newblock {Minimal tensors and purely electric or magnetic spacetimes of
  arbitrary dimension}.
\newblock {\em Classical and Quantum Gravity}, 30:165014, 2013.
\newblock \href{http://arxiv.org/abs/1203.3563}{{arXiv:1203.3563 [gr-qc]}}.

\bibitem{Mac86}
M.~A.~H. MacCallum.
\newblock Computer-aided classification of exact solutions in general
  relativity.
\newblock In H.~Sato and T.~Nakamura, editors, {\em Gravitational Collapse and
  Relativity (Proceedings of the XIV Yamada conference)}, pages 127--140. World
  Scientific, Singapore, 1986.

\bibitem{Olverbook}
P.~J. Olver.
\newblock {\em Equivalence, invariants, and symmetry}.
\newblock Cambridge University Press, Cambridge, 1995.

\bibitem{MilWyl13}
R.~Milson and L.~Wylleman.
\newblock {Three-dimensional spacetimes of maximal order}.
\newblock {\em Classical and Quantum Gravity}, 30:095004, 2013.
\newblock \href{http://arxiv.org/abs/1210.6920}{{arXiv:1210.6920 [gr-qc]}}.

\bibitem{MacAma86}
M.~A.~H. MacCallum and J.~E. {\AA}man.
\newblock Algebraically independent $n$-th derivatives of the {Riemannian}
  curvature spinor in a general spacetime.
\newblock {\em Classical and Quantum Gravity}, 3(6):1133--41, 1986.

\bibitem{ref4}
R.~Milson, A.~Coley, V.~Pravda, and A.~Pravdov\'a.
\newblock Alignment and algebraically special tensors in lorentzian geometry.
\newblock {\em International Journal of Geometric Methods in Modern Physics},
  2:41, 2005.
\newblock \href{http://arxiv.org/abs/gr-qc/0401010}{{arXiv:gr-qc/0401010
  [gr-qc]}}.

\bibitem{OPP2011}
M.~Ortaggio, V.~Pravda, and A.~Pravdov\'a.
\newblock Algebraic classification of higher-dimensional spacetimes based on
  null alignment.
\newblock {\em Classical and Quantum Gravity}, 30(1):013001, 2011.
\newblock \href{http://arxiv.org/abs/1211.7289}{{arXiv:1211.7289 [gr-qc]}}.

\bibitem{Hervik:2010rg}
S.~Hervik and A.~Coley.
\newblock {Curvature operators and scalar curvature invariants}.
\newblock {\em {Classical and Quantum Gravity}}, 27:095014, 2010.
\newblock \href{http://arxiv.org/abs/1002.0505}{{arXiv:1002.0505 [gr-qc]}}.

\bibitem{GoodeWain}
S.~W. Goode and J.~Wainwright.
\newblock {Characterization of locally rotationally symmetric space-times}.
\newblock {\em General Relativity and Gravitation}, 18:315--331, 1986.

\bibitem{Bradley1986}
M.~Bradley.
\newblock Construction and invariant classification of perfect fluids in
  general relativity.
\newblock {\em Classical and Quantum Gravity}, 3(3):317--334, 1986.

\bibitem{Seixas1992b}
W.~Seixas.
\newblock {Killing vectors in conformally flat perfect fluids via invariant
  classification}.
\newblock {\em Classical and Quantum Gravity}, 9(1):225, 1992.

\bibitem{WylCosNat19}
L.~Wylleman, F.~Costa, and J.~Natario.
\newblock Poynting vector, super-poynting vector, and principal observers in
  electromagnetism and general relativity.
\newblock {\em ArXiv e-prints}, 2020.
\newblock \href{http://arxiv.org/abs/gr-qc/2007.15384}{{arXiv:2007.15384
  [gr-qc]}}.

\bibitem{WylVdB06}
L.~Wylleman and N.~Van~den Bergh.
\newblock {Complete classification of purely magnetic, non-rotating and
  non-accelerating perfect fluids}.
\newblock {\em Physical Review D}, 74:084001, 2006.
\newblock \href{http://arxiv.org/abs/gr-qc/0604025}{{arXiv:gr-qc/0604025
  [gr-qc]}}.

\bibitem{mcintosh1994electric}
C.~B.~G. McIntosh, R.~Arianrhod, S.~T. Wade, and C.~Hoenselaers.
\newblock {Electric and magnetic Weyl tensors: classification and analysis}.
\newblock {\em Classical and Quantum Gravity}, 11(6):1555, 1994.

\bibitem{Visser2018}
P.~Martin-Moruno and M.~Visser.
\newblock {Essential core of the Hawking--Ellis types}.
\newblock {\em Classical and Quantum Gravity}, 35(12):125003, 2018.
\newblock \href{http://arxiv.org/abs/1802.00865}{{arXiv:1802.00865 [gr-qc]}}.

\bibitem{CHDG}
A.~A. Coley, S.~Hervik, M.~N. Durkee, and M.~Godazgar.
\newblock Algebraic classification of five-dimensional spacetimes using scalar
  invariants.
\newblock {\em Classical and Quantum Gravity}, 28(15):155016, 2011.
\newblock \href{http://arxiv.org/abs/1105.2355}{{arXiv:1105.2355 [gr-qc]}}.

\bibitem{CH}
A.~{Coley} and S.~{Hervik}.
\newblock {Algebraic classification of spacetimes using discriminating scalar
  curvature invariants}.
\newblock {\em ArXiv e-prints}, 2010.
\newblock \href{http://arxiv.org/abs/1011.2175}{{arXiv:1011.2175 [gr-qc]}}.

\bibitem{szekeres1975class}
P.~Szekeres.
\newblock A class of inhomogeneous cosmological models.
\newblock {\em Communications in Mathematical Physics}, 41(1):55--64, 1975.

\bibitem{tolman1934effect}
R.~C. Tolman.
\newblock {Effect of inhomogeneity on cosmological models}.
\newblock {\em Proceedings of the national academy of sciences of the United
  States of America}, 20(3):169, 1934.

\bibitem{StewartEllis}
J.~M. Stewart and G.~F.~R. Ellis.
\newblock {Solutions of Einstein's equations for a perfect fluid which exhibit
  local rotational symmetry}.
\newblock {\em Journal of Mathematical Physics}, 9(1):1072, 1968.

\bibitem{nolan2007shell}
B.~C. Nolan and U.~Debnath.
\newblock {Is the shell-focusing singularity of Szekeres space-time visible?}
\newblock {\em Physical Review D}, 76(10):104046, 2007.

\bibitem{georg2017symmetry}
I.~Georg and C.~Hellaby.
\newblock {Symmetry and equivalence in Szekeres models}.
\newblock {\em Physical Review D}, 95(12):124016, 2017.

\bibitem{bonnor1977rotating}
W.~B. Bonnor.
\newblock A rotating dust cloud in general relativity.
\newblock {\em Journal of Physics A: Mathematical and General}, 10(10):1673,
  1977.

\bibitem{Sussman}
R.~A. Sussman and K.~Bolejko.
\newblock {A novel approach to the dynamics of Szekeres dust models}.
\newblock {\em Classical and Quantum Gravity}, 29(6):065018, 2012.
\newblock \href{http://arxiv.org/abs/1109.1178}{{arXiv:1109.1178 [gr-qc]}}.

\bibitem{KS1966}
R.~Kantowski and R.~K. Sachs.
\newblock Some spatially homogeneous anisotropic relativistic cosmological
  models.
\newblock {\em Journal of Mathematical Physics}, 7(3):443--446, 1966.

\bibitem{kasner1921geometrical}
E.~Kasner.
\newblock {Geometrical theorems on Einstein's cosmological equations}.
\newblock {\em American Journal of Mathematics}, 43(4):217--221, 1921.

\bibitem{Hewitt:1991hf}
C.~G. Hewitt.
\newblock {An exact tilted Bianchi II cosmology}.
\newblock {\em Classical and Quantum Gravity}, 8(5):L109, 1991.

\bibitem{SM}
M.~Mars and J.~Senovilla.
\newblock Non-diagonal $\mathcal{G}_{2}$ separable perfect-fluid spacetimes.
\newblock {\em Classical and Quantum Gravity}, 14:205, 01 1997.

\bibitem{Harko:2002hv}
T.~Harko and M.~K. Mak.
\newblock {Viscous Bianchi type I universes in brane cosmology}.
\newblock {\em Class. Quant. Grav.}, 20:407--422, 2003.
\newblock \href{http://arxiv.org/abs/gr-qc/gr-qc/0212075}{{arXiv:gr-qc/0212075
  [gr-qc]}}.

\bibitem{Quevedo:2002xw}
F.~Quevedo.
\newblock {Lectures on string/brane cosmology}.
\newblock {\em {Classical and Quantum Gravity}}, 19:5721--5779, 2002.
\newblock \href{http://arxiv.org/abs/hep-th/0210292}{{arXiv:hep-th/0210292
  [hep-th]}}.

\bibitem{MilPel09}
R.~Milson and N.~Pelavas.
\newblock {The curvature homogeneity bound for Lorentzian four-manifolds}.
\newblock {\em International Journal of Geometric Methods in Modern Physics},
  6(01):99--127, 2009.
\newblock \href{http://arxiv.org/abs/0711.3851}{{arXiv:0711.3851 [gr-qc]}}.

\end{thebibliography}

\end{document}